# Generative AI for Requirements Engineering: A Systematic Literature Review

**Haowei Cheng**[1] | **Jati H. Husen**[1,2] | **Yijun Lu**[1] | **Teeradaj Racharak**[3,4] | **Nobukazu Yoshioka**[1] | **Naoyasu Ubayashi**[1] | **Hironori Washizaki**[1]

[1]Waseda University, Tokyo, Japan

[2]Telkom University, Jawa Barat, Indonesia

[3] Advanced Institute of So-Go-Chi (Convergence Knowledge) Informatics, Tohoku University, Miyagi, Japan

[4] Japan Advanced Institute of Science and Technology (JAIST), Ishikawa, Japan

**Correspondence**
Haowei Cheng
Email: haowei.cheng@fuji.waseda.jp

## Abstract

**Introduction:** Requirements engineering (RE) faces challenges due to the handling of increasingly complex software systems. These challenges can be addressed using generative artificial intelligence (GenAI). Given that GenAI-based RE has not been systematically analyzed in detail, this review examines the related research, focusing on trends, methodologies, challenges, and future work directions.

**Methods:** A systematic methodology for paper selection, data extraction, and feature analysis is used to comprehensively review 238 articles published in 2019–2025 and available from major academic databases.

**Results:** Although generative pretrained transformer models dominate current applications (67.3% of studies), the research focus remains unevenly distributed across RE phases, with analysis (30.0%) and elicitation (22.1%) receiving the most attention and management (6.8%) remaining underexplored. Three core challenges—reproducibility (66.8%), hallucinations (63.4%), and interpretability (57.1%)—form a tightly interlinked triad affecting trust and consistency, and strong correlations (≥35% cooccurrence) indicate that these challenges must be addressed holistically. Industrial adoption remains nascent, with >90% of studies corresponding to early-stage development and only 1.3% reaching production-level integration. Evaluation practices show maturity gaps, limited tool/dataset availability, and fragmented benchmarking approaches.

**Conclusions:** Despite the transformative potential of GenAI-based RE, several barriers hinder its practical adoption. The strong correlations among core challenges demand specialized architectures targeting interdependencies rather than isolated solutions. The limited real-world deployment reflects systemic bottlenecks in generalizability, data quality, and scalable evaluation methods. Successful adoption requires coordinated development across technical robustness, methodological maturity, and governance integration. A multiphase research roadmap emphasizing evaluation infrastructure strengthening, governance-aware development, and industrial-scale standardization is proposed.

## 1 | INTRODUCTION

### 1.1 | Background and Motivation

Requirements engineering (RE) faces mounting challenges in today's rapidly evolving software landscape. With the increasing complexity and

**Abbreviations:** RE, requirements engineering; SE, software engineering; GenAI, generative AI.

interconnectivity of software systems, accurate requirement identification and management are becoming increasingly critical and problematic.

The primary contributors to cost overruns in software projects are frequent requirement changes and the lacking understanding of system need among stakeholders (Lederer and Prasad (1995)). The Standish Group's "2020 CHAOS Report" has further substantiated this observation. The report has revealed that merely 31% of software projects are completed within given time and budget constraints, and only 46% deliver high-value returns to stakeholders (Group (2020)). Thus, ineffective RE practices remain a major bottleneck of software engineering (SE) development, which underscores the criticality of innovative approaches in enhancing RE practices. Recent industry analyses have quantified the challenges in RE. According to a technical report by the Project Management Institute, failures resulting from requirement-related issues account for approximately 37% of project failures in enterprise software development, with an average cost impact of 25%–40% of the initial project budget (Project Management Institute (2014)). A comprehensive survey by Carnegie Mellon's Software Engineering Institute reveals that 68% of requirement defects are discovered only at the later stages of development or after deployment, wherein the cost of correction is 5–10 times higher than that during the requirements phase (Software Engineering Institute (2023)). These statistics underscore the criticality of more effective RE approaches, particularly for addressing the increasing complexity of modern software systems.

The integration of artificial intelligence (AI) into SE has transformed traditional practices, particularly tasks such as code generation, defect prediction, and software testing, which has substantially improving efficiency and quality across development processes (Svyatkovskiy et al. (2020), Chen et al. (2018), Motwani et al. (2018)). The ability of AI to automate and enhance these tasks has paved the way for further intelligent and adaptive SE practices. Building on these advancements, current research is increasingly focusing on extending the advantages of AI to RE, with the objective of addressing unique challenges in eliciting, analyzing, and validating software requirements. While previous studies have reviewed the applications of AI and machine learning (ML) in RE (e.g., Zamani et al., 2021 Zamani et al. (2021a)), these reviews have primarily concentrated on traditional ML techniques. With the emergence of large language models (LLMs), such as ChatGPT, the use of generative AI (GenAI) in RE provides new opportunities but faces certain problems. To the best of our knowledge, this systematic literature review (SLR) is the first to focus on the use of GenAI in RE, not only extending the scope of existing reviews on AI applications in RE but also focusing on the unique characteristics of GenAI, such as prompt engineering and hallucination issues, and their implications in RE. This shift marks an important evolution from the traditional role of AI in SE to the GenAI-based optimization of RE practices. The AI-based approaches used in SE lay the foundation for the application of more advanced (GenAI) techniques such as LLMs in various SE tasks. The success of AI in improving software development processes and quality sets the stage for exploring the potential of GenAI in RE. To understand the manner in which GenAI can enhance software development processes, the fundamental components of RE should first be examined.

### 1.2 | Contribution and Paper Organization

This SLR makes three interconnected contributions to GenAI applications in RE, reviewing and critically analyzing current research by examining 238 papers published between 2019 and May 2025. A multidimensional classification framework enabling the systematic evaluation of GenAI–RE research across publication metadata, technical configurations, prompt engineering strategies, evaluation practices, and industrial adoption statuses is used to support this analysis, and a research roadmap providing insights for researchers and practitioners is derived.

Section 2 presents related work and context on the use of AI in RE, GenAI in SE, and GenAI in RE. Section 3 outlines the research questions (RQs) guiding this SLR. Section 4 details the research methodology, including the systematic search strategy, rigorous data inclusion/exclusion criteria, and data extraction process. Section 5 presents the results, analysis, and discussion organized around the RQs, examining the current research trends; predominant approaches, techniques, and quality objectives; evaluation methods, metrics, tools, and dataset support; major challenges and their future implications; and industrial adoption statuses and trends. Section 6 examines threats to validity and discusses the related mitigation strategies. Section 7 presents a roadmap for advancing GenAI applications in RE based on our findings. Section 8 summarizes the key findings and provides insights for future research on GenAI applications in RE.

## 2 | RELATED WORK

The application of AI to SE has emerged as a rapidly evolving field, with recent advancements in GenAI opening new frontiers and reshaping traditional practices. This section provides a comprehensive overview of the existing research landscape, focusing on key areas that relate to our research: the evolution from traditional AI in SE to specialized applications in RE, and the emergence of GenAI across both SE and RE domains.

Traditional AI approaches have been successfully used in SE across various software development tasks. Several SLRs have summarized this state-of-the-art, with comprehensive reviews covering ML techniques for defect prediction, effort estimation, and code generation Brar and Nandal (2022), Tian et al. (2021). These AI-based approaches in SE establish the foundation for more advanced GenAI techniques and demonstrate the evolution toward the use of current generative models in RE.

Building on this foundation, we examine three key research areas, namely the application of AI in RE, broad impact of GenAI use in SE, and the emerging use of GenAI in RE.

## 2.1 | Requirements Engineering Overview

RE plays a critical role in SE because it focuses on the systematic elicitation, analysis, specification, validation, and management of functional and nonfunctional requirements (Nuseibeh and Easterbrook (2000)). The importance of RE in software development has been widely recognized and codified in international standards such as ISO/IEC/IEEE 29148:2018. These standards provide a unified framework and best-practice guidelines for the RE process, which emphasizes the crucial role of RE in project success (Int (2018)).

As shown in Figure 1, RE encompasses five primary components:

1. **Elicitation:** Diverse methods, e.g., structured interviews, collaborative workshops, and systematic surveys, are used to elicit stakeholder needs.
2. **Analysis:** Modeling techniques and prioritization frameworks are used based on importance and implementation feasibility.
3. **Specification:** Requirements are documented in various formats, including user stories, use cases, and formal specifications, to ensure the clear communication of system expectations.
4. **Validation:** Requirement accuracy is ensured using rigorous review processes, prototyping, and testing methodologies.
5. **Management:** This overarching component involves change control, traceability, and version control, ensuring that the requirements remain consistent and up-to-date throughout the project lifecycle.

Traditional RE methods typically face challenges related to efficiency and accuracy, especially when handling rapidly evolving and increasingly complex requirements. As modern software systems increase in terms of scale and intricacy, enhancing the quality and effectiveness of RE processes remains a pressing issue in contemporary SE (Wiegers and Beatty (2013)).

Several distinct phases, which are driven by the need to overcome increasingly complex software development challenges, have marked the evolution of RE (Nuseibeh and Easterbrook (2000)). Traditional RE methods heavily relied on manual processes, stakeholder interviews, and document-centric approaches. As software systems became increasingly intricate, these methods have proved ineffective in capturing and managing the full spectrum of requirements. This has led to the rise of model-driven RE, which incorporates visual representations and formal specifications to improve clarity and traceability (Cheng and Atlee (2007)). In response to the increasing complexity and dynamic nature of modern software systems, RE has gradually evolved into AI-driven approaches that automate many repetitive tasks, improve analysis capabilities, and handle larger volumes of data with greater efficiency (Franch (2021)). Recently, this evolution has taken another step with the introduction of GenAI into RE. By leveraging advanced technologies, such as LLMs, GenAI for RE represents a new paradigm that not only improves traditional RE activities but also elicits new possibilities for the automation of the generation, refinement, and analysis of requirements (Zamani et al. (2021b), Marques et al. (2024a)). This shift toward GenAI-for-RE marks a transformative point in the RE field, which offers unprecedented potential for handling the increasing complexity of requirements and evolving stakeholder needs. As the exploration of how to overcome challenges in modern RE continues, GenAI emerges as a promising solution that offers unique capabilities and approaches.

## 2.2 | GenAI and Its Components

GenAI represents a fundamental shift in computational intelligence and is characterized by the ability to generate novel and coherent content across multiple modalities, including text, images, code, and structured data Shneiderman (2020). Unlike earlier AI systems limited to classification or retrieval tasks, GenAI systems are designed to synthesize content that aligns with human-like contextual relevance, linguistic fluency, and domain conventions.

Three interdependent components, namely deep neural architectures, self-supervised learning algorithms, and large-scale pretraining data, lie at the core of GenAI Vaswani et al. (2017). Transformer-based models, exemplified by generative pretrained transformer (GPT)-4, enable the encoding of long-range dependencies via attention mechanisms, facilitating context-aware text generation and crosslingual reasoning. These architectures dynamically adjust the focus across input sequences, yielding outputs that are semantically precise and syntactically coherent.

The learning algorithms employed by GenAI systems combine supervised and self-supervised techniques, placing particular emphasis on masked language modeling and autoregressive objectives. Self-supervised learning allows models to acquire generalized language representations without requiring labeled data and enable transferability across RE domains. Optimization strategies such as curriculum learning and few-shot adaptation further enhance the capability of the model to generalize and specialize simultaneously.

Finally, the performance of GenAI systems critically depends on the scale and diversity of pretraining data. Vast corpora spanning technical documentation, natural language dialogues, and domain-specific artifacts serve as the foundation upon which GenAI models learn to emulate human-like output. This triadic foundation—architecture, algorithm, and data—underpins the remarkable performance of modern GenAI in tasks ranging from conversational reasoning to structured requirement generation.

## 2.3 | AI in RE

In the RE domain, AI has emerged as a promising solution for automating and enhancing various RE tasks, which addressed challenges associated with the increased complexity and scale of modern software systems. Talele et al. proposed an innovative AI-driven approach for requirements classification and prioritization, which significantly reduced the manual effort required in these critical processes (Talele and Phalnikar (2021)). This work demonstrated the potential of AI in streamlining requirement management and enabling effective decision-making in software development projects. In addition, studies such as Rahimi et al. (Rahimi et al. (2014)) have successfully used AI techniques in requirements traceability, which showcased improved accuracy in linking requirements to other software artifacts such as design documents and test cases. These advancements have highlighted the transformative potential of AI in optimizing RE activities, ensuring requirement consistency, and facilitating effective communication among stakeholders. By leveraging AI, researchers and practitioners can unlock new possibilities for automating and streamlining RE processes, which ultimately leads to high-quality software systems that better align with user needs and expectations.

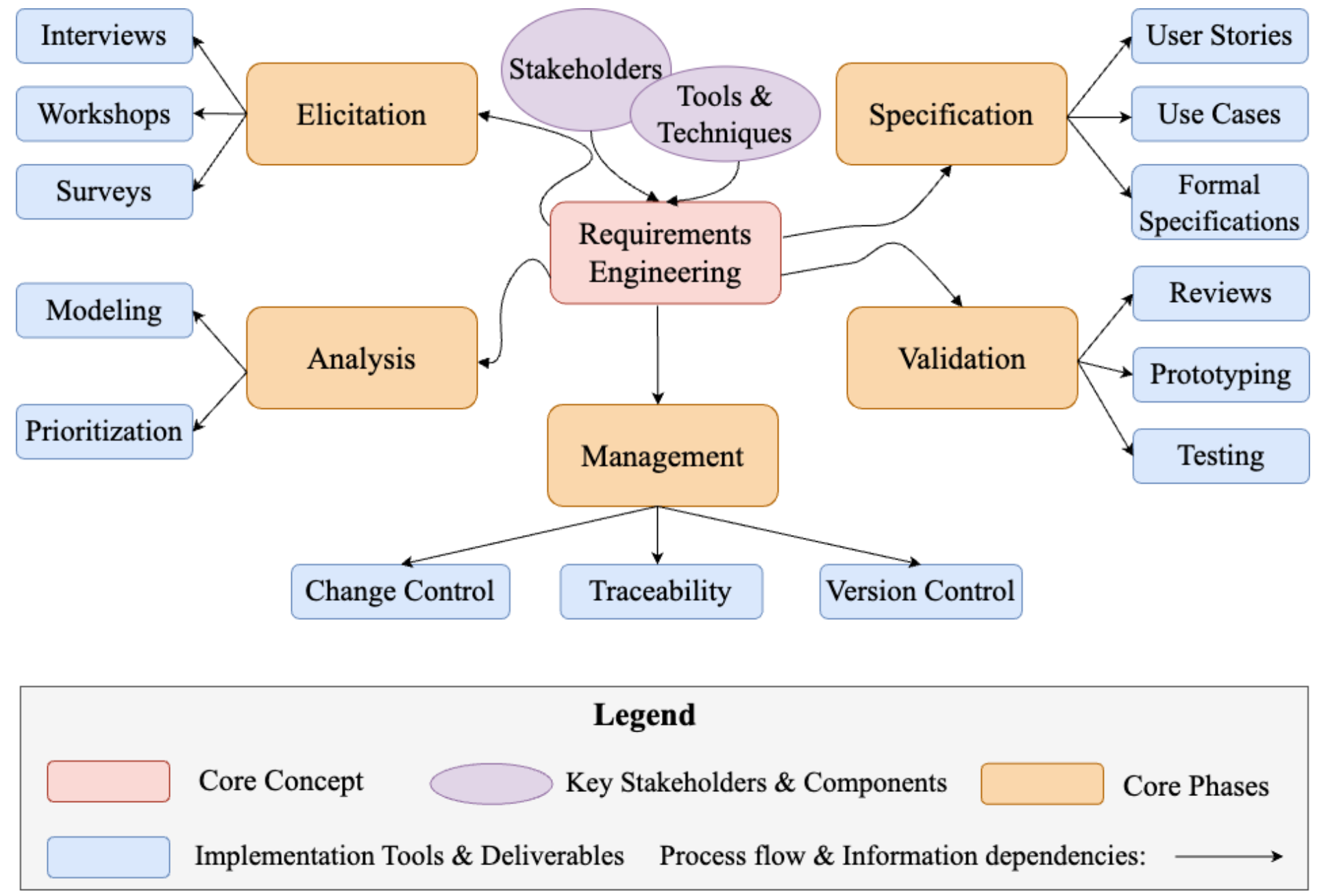

**FIGURE 1** Concept of requirements engineering (RE).

Zamani et al.'s comprehensive mapping study of ML in RE provided a milestone in understanding the landscape of AI applications in RE (Zamani et al. (2021a)). Their systematic analysis of RE tasks supported by ML revealed the extent of potential applications and challenges in transitioning from laboratory experiments to industrial practice. This work established a foundational understanding of the effective integration of ML techniques into various RE activities, while highlighting the need for other sophisticated approaches to address complex challenges in RE. Necul et al. (Necula et al. (2024)) conducted an SLR on NLP in RE by examining requirements classification, information extraction, and quality assessment (QA), demonstrating the potential of NLP in automating RE tasks while highlighting challenges in terms of the lack of RE-specific tools and domain knowledge requirements. Zhao et al. conducted a comprehensive systematic mapping study and reviewed 404 primary studies in NLP for RE across 36 years. The results revealed that while numerous NLP technologies have been developed, numerous studies remain at the laboratory level with limited industrial validation. Their study highlighted that requirements analysis is the most addressed RE phase with a central focus on the detection of quality defects (Zhao et al. (2021)).

Future research directions in AI for RE should focus on addressing these challenges. A pressing need exists to develop AI tools specific to RE that can better handle the complexities of requirements elicitation, analysis, and management. In addition, effort should be directed toward the enhancement of the domain adaptability of AI models, which enables them to effectively process and interpret domain-specific requirements across industries. Improving the explainability and transparency of AI-generated outputs in RE is another critical area for future work, because it directly influences the trustworthiness and adoption of these technologies in practice.

The application of AI techniques in RE has demonstrated potential in automating and enhancing various RE tasks. These advancements have paved the way for the exploration of more sophisticated GenAI models, which could further revolutionize the elicitation, analysis, and management of requirements.

## 2.4 | GenAI in SE

The emergence of GenAI models, particularly LLMs, has ushered in a new era of possibilities in SE, which has enabled the automated generation of code, documentation, and other software artifacts. Mastropaolo et al. conducted a groundbreaking study on the use of the text-to-text transfer transformer to support code-related tasks, which highlighted the immense potential of generative models in enhancing software development processes (Mastropaolo et al. (2021)). This work paved the way for more intelligent and efficient code generation techniques, thereby reducing the manual effort required for software development. Similarly, Fried et al. introduced InCoder, a state-of-the-art generative model for code infilling and synthesis, and demonstrated its effectiveness in generating code snippets based on natural language descriptions (Fried et al. (2023)). This innovative approach showcased the potential of GenAI in enabling other natural and intuitive ways of expressing software requirements and specifications. However, challenges remain in ensuring the reliability and security of AI-generated code, as highlighted by Zong et al. (Zong et al. (2019)). These studies underscored the immense potential of GenAI in automating and augmenting various SE tasks while emphasizing the need for further research to address related challenges and ensure the trustworthiness and robustness of GenAI-assisted software development.

A comprehensive research agenda by Nguyen-Duc et al. (Nguyen-Duc et al. (2023)) identified 78 open RQs across 11 areas of SE to which GenAI can be applied. Their agenda covers a wide range of topics, including RE, software design, implementation, quality assurance, maintenance, processes, project management, professional competencies, education, macro aspects, and the fundamental concerns of GenAI in SE. Xu et al. (Xu et al. (2024)) explored novel program synthesis approaches by combining LLMs with automated ML, demonstrating promising results in automating entire ML workflows from task descriptions. This research agenda highlights the need for further research to address the challenges associated with GenAI-assisted software development. These challenges include industry-level evaluation, reliability and correctness concerns, data availability, explainability, and sustainability aspects. The agenda serves as a valuable resource for researchers and practitioners to guide efforts in future research and development in this rapidly evolving field.

Another SLR by Li et al. (Jiang et al. (2024)) specifically focused on the use of LLMs for code generation. The review introduced a taxonomy to categorize and discuss recent developments in code LLMs, which covers aspects such as data curation, latest advances, performance evaluation, and real-world applications. The authors provided a historical overview of the evolution of LLMs for code generation and offered an empirical comparison that used widely recognized benchmarks to highlight the progressive enhancements in the capabilities of LLMs. They identified critical challenges and promising opportunities for the gap between academia and practical development, which emphasized the need for a dedicated resource to continuously document and disseminate the most recent advances in the field. Similarly, systematic investigations have also been conducted to understand the benefits and challenges of integrating GenAI into self-adaptive systems, providing insights into both technical capabilities and practical implementation considerations (Li et al. (2024)).

The current study builds on these existing reviews and specifically focuses on the application of GenAI techniques in the context of RE. The successful application of GenAI techniques in various SE tasks, such as code generation and documentation, highlights the potential application of similar approaches in the RE domain. The lessons learned and challenges identified in GenAI for SE provide valuable insights for researchers and practitioners that are exploring the use of GenAI in RE.

**Comparison with existing SE surveys:**

Among the existing SE-wide studies, Hou et al. (2023) conducted an SLR following a well-defined protocol, covering 395 primary studies published in 2017-2024. The authors systematically mapped LLM-based SE applications to various SE activities and summarized research trends, challenges (e.g., hallucination and reliability), and technical strategies (e.g., prompt engineering and hybrid techniques). However, RE treated as a minor category, and no detailed synthesis of RE-specific techniques, challenges, or industrial adoption was provided.

Fan et al. (2023) surveyed the use of LLMs in SE, highlighting their potential in code generation, testing, debugging, maintenance, and documentation while noting challenges such as hallucination, correctness, non-determinism, and security. The author emphasized the importance of hybrid approaches combining LLMs with traditional SE techniques, especially automated testing and search-based SE for reliability and efficiency. Open research directions, including SE-specific LLM training, dynamic prompt optimization, rigorous evaluation, and even the positive leveraging of hallucinations, were also outlined to provide a roadmap for future LLM-driven SE research.

Unlike these SE-wide works, the present study focuses exclusively on the applications of GenAI in RE. By incorporating additional databases (ACM DL and Google Scholar), applying snowballing techniques, and extending the search period to May 2025, our work provides a fine-grained and up-to-date synthesis of RE-specific techniques, challenges, and practical implications, which are largely missing from SE-wide surveys.

## 2.5 | GenAI in RE

By examining these interconnected domains, the current SLR on GenAI for RE is strategically positioned at the convergence of AI, SE, and RE. While comprehensive reviews exist on AI applications in SE (Brar and Nandal (2022), Tian et al. (2021)), AI in RE (Necula et al. (2024), Zhao et al. (2021)), and GenAI in SE (Nguyen-Duc et al. (2023), Jiang et al. (2024)), systematic reviews that emphasize GenAI applications in RE remain scarce. Although Marques et al. (Marques et al. (2024b)) reported an initial exploration that specifically examined ChatGPT in RE, a comprehensive review of broader GenAI applications in RE remains lacking. The current review aims to fill this research gap by building on collective knowledge in related domains to offer a comprehensive analysis of the current landscape, challenges, and future trajectories of GenAI applications in RE.

The application of GenAI in RE is a nascent field with an immense potential to transform the manner in which software requirements are elicited, analyzed, and validated. Santos et al. recent study investigated the use of GenAI to automatically generate design practices and evaluate their ability to satisfy specific requirements (Santos et al. (2024)). This approach showcased the potential of GenAI in assisting requirements engineers in exploring and evaluating alternative design solutions, facilitating creative and efficient RE processes. Moreover, ChatGPT, a state-of-the-art LLM, has been used to automatically detect inconsistencies in natural language requirements, which enhanced the requirements validation process (Fantechi et al. (2023)). This work suggested that GenAI could assist in detecting inconsistencies in software requirements, aiding the requirements validation process.

Another important area of research explores the potential of ChatGPT in assisting in requirements elicitation processes by evaluating the quality of requirements generated by a model and comparing them with those formulated by human RE experts (Ronanki et al. (2023)). This comparative analysis provides valuable insights into the strengths and limitations of GenAI in capturing and expressing user requirements and expectations, which paves the way for more effective human–AI collaboration in RE. In a recent review, Marques et al. (Marques et al. (2024b)) focused on the use of ChatGPT in RE. They discussed the potential applications of ChatGPT in requirements elicitation, analysis, and validation, including its limitations and challenges, such as the need for human oversight and the potential for biased or inconsistent output. The current review builds on this growing body of research, with the objective of systematically analyzing the current state of GenAI applications in RE and providing a roadmap for future research and development in this field.

Our analysis of challenges faced by LLMs used in RE was guided by a comprehensive framework proposed by Naveed et al. (Naveed et al. (2024)). Their survey identified critical challenges, including those related to computational costs, bias and fairness, interpretability, and safety. These challenges are particularly relevant to RE, wherein biased or incorrect requirements can lead to discriminatory or unreliable systems. The emphasis of the framework on hallucinations and ethical deployment helped in the development of our approach toward ensuring reliable and accurate outputs across diverse RE projects.

For instance, our discussion on bias and fairness was informed by the emphasis placed by Naveed et al. (Naveed et al. (2024)) on the potential societal implications of these challenges. The issue of hallucinations in LLMs, which is a key point of their survey, guided our approach for addressing the challenge of generating incorrect or inconsistent requirements in the context of RE, in which models need to provide reliable and accurate outputs across diverse projects and domains. Furthermore, our examination of challenges in safety and controllability was enriched by the perspectives presented in their work, particularly in relation to the ethical deployment of LLMs. This aspect is especially crucial in RE, wherein model outputs can directly influence critical system specifications.

The alignment of our approach with this established framework facilitates a comprehensive and systematic examination of the multifaceted nature of LLMs in the context of RE. Section 5 presents a detailed analysis of the dimensions based on the findings of our SLR. The analysis encompasses specific research trends, technological methods employed, QA, and future directions in this domain, with a particular focus on the application of models such as GPT-3 and GPT-4 in various RE stages and how techniques, including prompt engineering and few-shot learning, enhance model performance in RE tasks. This approach is essential for realizing the full potential of GenAI in revolutionizing RE processes and achieving more efficient, accurate, and ethically sound software development practices.

**Comparison with existing RE-focused SLRs:**

The most closely related study is the recent SLR by Khan et al. (2025), who reviewed 35 studies on LLM applications in RE published in 2023-2024. This work mainly focused on classifying RE activities (elicitation, specification, modeling, validation, prioritization, and traceability) and summarized prompting strategies, evaluation metrics, and generic challenges.

Compared with that of Khan et al. (2025), the present study presents several notable differences. First, we expanded the database coverage by including ACM DL, Google Scholar, and ArXiv and applied both backward and forward snowballing to ensure maximum coverage. Second, the search was extended to May 2025 to capture the most recent LLM-based RE studies, including those dealing with emerging models such as CodeLlama and Gemini. Third, beyond basic categorization, we provided deeper analyses, such as cross-analysis between LLM types and RE-specific challenges, as well as prompting effectiveness comparisons. Fourth, we integrated industrial adoption trends, dataset and tool availability, and practical implications into a GenAI–RE framework, providing guidance for researchers and practitioners. These aspects are largely absent in the work of Khan et al.'s work, which primarily focuses on descriptive statistics.

## 2.6 | GenAI-RE Integration

Figure 2 presents a conceptual view of how GenAI technologies can support and enhance RE activities. This view is structured along three key dimensions constructed by combining findings from existing works on GenAI capabilities and RE processes. The visualization highlights the layered relationships between GenAI functionalities and RE practices, with directional flows representing the functional progression from technological potential to practical outcomes.

The architecture comprises three interconnected conceptual dimensions, namely GenAI capabilities (top), a translation layer (middle), and the RE lifecycle (bottom). The top and bottom layers represent the source technologies and target application phases, respectively. The translation layer serves as a bridge, capturing intermediate effects that allow GenAI to be effectively applied in RE. These effects include enhanced requirement quality, operational efficiency, and traceability.

The capability layer includes four commonly discussed GenAI functionalities, namely NLP, pattern recognition, predictive analytics, and automated documentation. These functionalities are typically realized through LLMs, which offer context understanding, content generation, and cross-lingual reasoning capabilities. The translation layer captures the way in which these core functions enable measurable impacts within RE practices.

In the elicitation phase, the NLP capabilities of GenAI improve stakeholder communication by generating clarifying questions and uncovering implicit requirements. Pattern recognition techniques analyze past project data to uncover common requirement patterns and detect omissions. During specification, automated documentation supports the creation of consistent, structured drafts. Validation leverages predictive analytics to detect potential conflicts, ambiguities, and completeness gaps, which is especially valuable in large-scale projects. For requirement management, GenAI supports change impact analysis and traceability, facilitating effective requirement evolution over time.

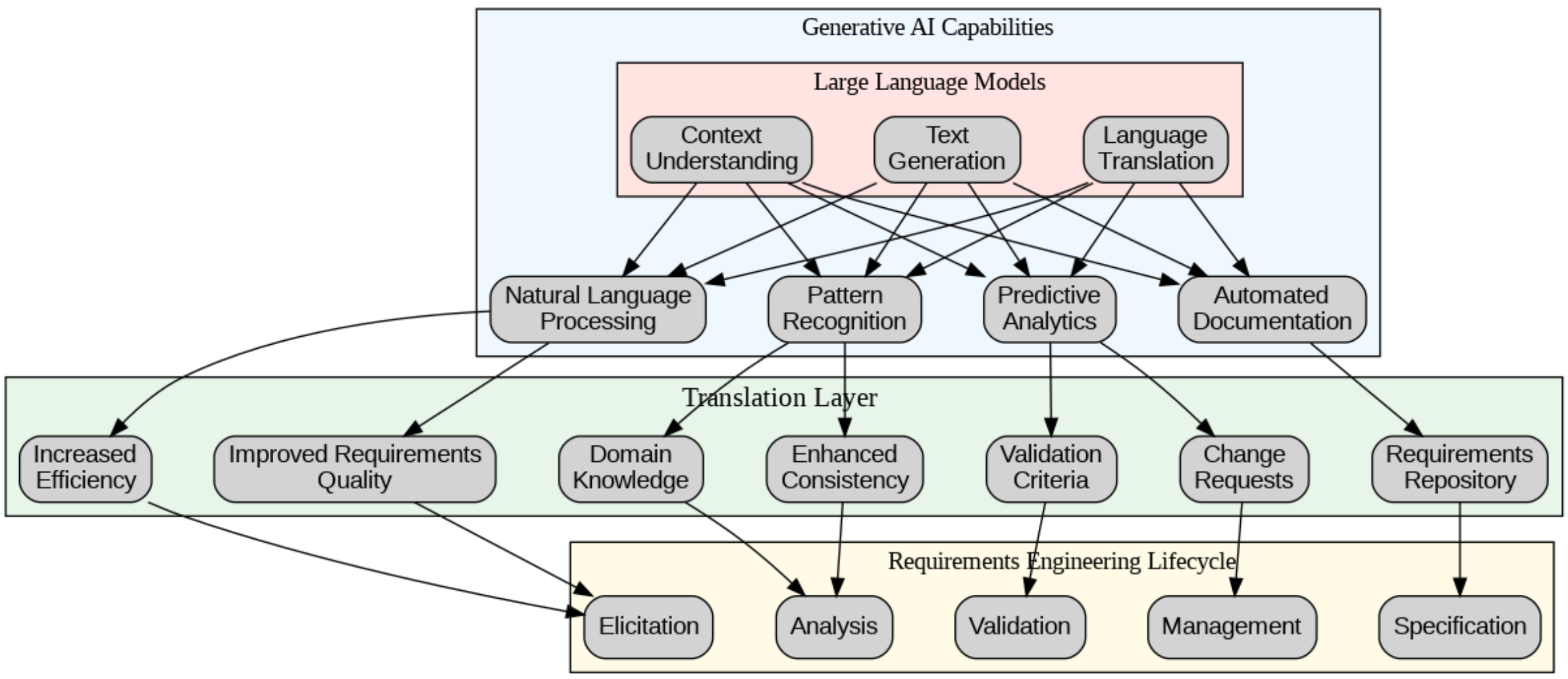

**FIGURE 2** Conceptual dimensions of generative artificial intelligence (GenAI) integration into RE. The top and bottom layers represent GenAI capabilities and RE lifecycle phases, respectively. The translation layer (middle) bridges these layers by capturing functional effects such as improved requirement quality, efficiency, and traceability.

To illustrate how GenAI capabilities translate into RE improvements, one can consider the following pathway. Context understanding, enabled by LLM-based NLP, facilitates stakeholder communication by generating clarifying questions and identifying implicit requirements during elicitation. These inputs are then processed through pattern recognition techniques to detect omissions and align with common requirement patterns, thereby enhancing specification consistency. Predictive analytics further support validation by identifying conflicts, ambiguities, and completeness issues, particularly in large-scale or time-sensitive projects.

Such translation chains are evident in recent studies. For example, Sarferaz (2025) integrated GenAI into a production-level ERP system, where LLMs facilitated change impact analysis and the automated validation of the evolving requirements, providing traceability and scalability benefits. Similarly, Ben Chaaben et al. (2024) proposed a graph-based requirement generation system using few-shot and chain-of-thought (CoT) prompting followed by structured documentation and case study-based evaluation, which well aligned with the full elicitation-to-validation cycle. Liu et al. (2025) used the Gensors tool to transform user intent into visualized personalized requirement drafts and then validated them through structured user studies to assess consistency and perceived control. These studies exemplify how GenAI capabilities, when mediated through translation effects such as enhanced quality or traceability, contribute to multiple phases of the RE lifecycle.

# 3 | RESEARCH QUESTIONS

RE is critically important for software development and involves requirements elicitation, analysis, specification, validation, and management. The increasing complexity of software systems and demand for efficient and effective RE processes have led to the exploration of advanced AI techniques, particularly GenAI, to support and enhance various RE activities.

Our SLR aims to provide a comprehensive overview of the current state of research on the application of GenAI techniques in RE, elucidating the key research trends, methodologies employed, and challenges encountered in this emerging field by analyzing the existing body of knowledge. Furthermore, we intended to provide insights into the potential of GenAI in addressing the limitations of traditional RE approaches and propose future research directions to advance the field. To systematically investigate the application of GenAI in RE, we formulated the following RQs:

**RQ1: What are the current research trends in the application of GenAI in RE?**

This RQ aims to analyze the distribution and characteristics of published studies, including the publication venues, temporal trends, and geographical distributions of the research efforts. By examining these aspects, we aim to understand the evolving landscape of GenAI applications in RE.

**RQ2: What are the current predominant GenAI approaches and techniques used in RE and their quality characteristics?**

This RQ examines the technical landscape and quality characteristics of GenAI applications used in RE. We review various technical approaches, including prompt engineering techniques, model architectures, fine-tuning strategies and LLM application strategy. Additionally, we investigate quality characteristics based on the ISO/IEC 25059 standard, as understanding quality characteristics is crucial for evaluating the practical effectiveness of GenAI solutions employed in RE tasks. This combined analysis helps identify current research focuses and potential gaps in terms of technicality and QA.

**RQ3:What are the evaluation methods, metrics, tools, and datasets supporting the use of LLMs in RE?**

This RQ investigates the evaluation landscape for GenAI applications in RE, examining tool and dataset availability, evaluation methodologies (experimental designs, case studies, user studies), performance metrics used (from traditional NLP metrics to RE-specific measures), and reproducibility practices. The abovementioned analysis helps assess the methodological maturity of the field and identify gaps in the evaluation infrastructure.

**RQ4: What are the main challenges of applying GenAI in RE, and they reflect the current research limitations? What future research directions can address these limitations?** This RQ addresses three crucial aspects of GenAI application in RE, identifying the primary challenges encountered in implementing GenAI in RE, exploreing potential future research trajectories in this rapidly evolving field, and examining how these challenges and future directions are related to the current limitations.

**RQ5: What is the current status and trend of the industrial adoption of LLMs in RE?**

This RQ investigates how LLMs are being adopted in industrial RE practices, examining the types of adoption (e.g., concept-level exploration, prototype development, or production-ready deployment), application scenarios and the industrial sectors involved. By analyzing these trends, we aim to identify the maturity level of GenAI-based RE solutions in practice, understand the existing gaps between academic research and industrial implementation, and highlight potential opportunities for industry–academia collaboration to accelerate practical adoption.

## 4 | RESEARCH METHODOLOGY

This SLR adheres to the widely recognized guidelines proposed by Kitchenham et al. (Keele et al. (2007)) for conducting systematic reviews in SE. These guidelines provide a rigorous framework for planning, conducting, and reporting the review process. Based on these guidelines, we developed a comprehensive protocol that includes systematic search strategies, explicit data inclusion/exclusion criteria, and comprehensive data extraction and synthesis procedures. To ensure a comprehensive and representative literature review, we employed a systematic search strategy for retrieving relevant publications. For analyzing the collected literature, we adopted the RE framework outlined in ISO/IEC/IEEE 29148:2018 (Int (2018)) as a reference point for classifying and evaluating the application of GenAI in RE.

To maximize coverage and ensure the comprehensive retrieval of relevant literature, we conducted searches across multiple academic databases. Scopus served as the main search engine because of its effectiveness in SLR and capability to export search results. According to the empirical study of (Baas et al. (2020)), Scopus provides high-quality bibliometric data with a comprehensive coverage of major publishers, including Elsevier, Springer Nature, Taylor & Francis, and Wiley. To maximize the retrieval of pertinent literature, the search was extended to ACM Digital Library, IEEE Xplore, Web of Science, Google Scholar, and ArXiv using identical query parameters. Although ArXiv is not typically used in literature reviews because of its non-peer-reviewed content, it was included because of the limited number of papers on the use of GenAI in RE available on Scopus. Despite the lack of peer review, ArXiv papers typically present innovative ideas and insights valuable for the current review. This multidatabase approach was adopted to minimize the risk of missing relevant publications and ensure broad coverage across different academic communities and publication venues.

The literature search was conducted on May 31, 2025, and covered works published between 2019 and May 2025 to capture the most recent advances. We selected the year 2019 as the starting point because of the emergence and development of GenAI technologies at this time, marked by the release of GPT-2, which is a milestone in the advancement of LLMs (Radford et al. (2019)).

The initial search yielded 772 papers. After meticulous screening, 238 papers were selected. The first author independently conducted screening, while the second and third authors assisted with analysis and validation. Disagreements were resolved through discussion and consensus. All search results and screening processes were systematically managed and documented using reference management software (Zotero). Figure 3 presents the paper selection and processing flows.

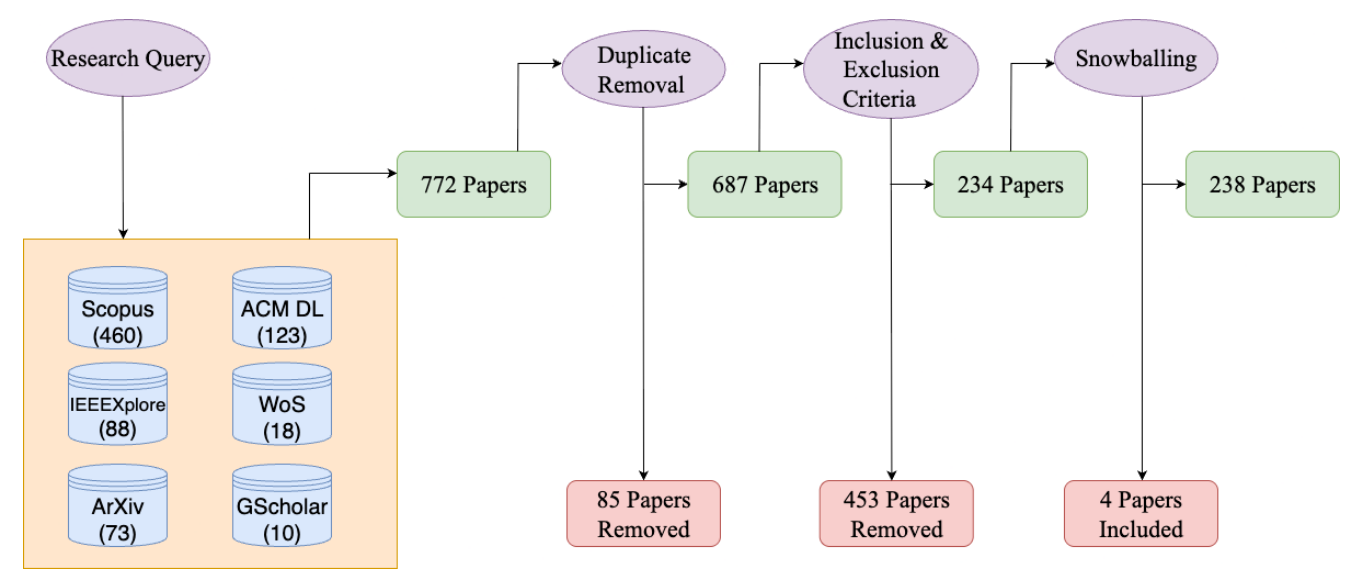

**FIGURE 3** Paper selection and processing flow

## 4.1 | Search and Selection Process

### 4.1.1 | Initial Search

The following query was executed on the titles, abstracts, and keywords of papers, with the publication period limited to 2019–May 2025:

```
("Generative AI" OR "Generative Artificial
Intelligence" OR "Large language model" OR "LLM" OR
"GPT") AND ("Requirements Engineering" OR "Requirements
Elicitation" OR "Requirements Analysis" OR "Requirements
Specification" OR "Requirements Validation" OR
"Requirements Management")
```

The search query was designed based on two major conceptual dimensions.

- Technology: We captured the core technology using "Generative AI" and "Generative Artificial Intelligence" as primary terms, along with specific technical implementations including "Large Language Model," "LLM," and "GPT." These keywords reflect the mainstream technical approaches used in GenAI research.
- Application domain: This dimension encompassed the RE domain. which was broken down into its major activities: elicitation, analysis, specification, validation, and management.

For each dimension, multiple related terms and synonyms were connected using OR operators to improve recall while maintaining precision.

### 4.1.2 | Omission of Duplicate Papers

Duplicate papers resulting from the overlap of search results across the Scopus, ACM Digital Library, IEEE Xplore, Web of Science, Google Scholar and ArXiv databases were omitted. The deduplication process eliminated 85 papers, which resulted in a final corpus of 687 publications.

### 4.1.3 | Data Inclusion and Exclusion Criteria

For paper selection, the first, second, and third (co)authors independently vetted the inclusion of each paper in the SLR using predefined criteria. Initially, we reviewed titles and abstracts, followed by full-text reading, to determine the relevance of each paper to GenAI and RE. Based on the query definition, 234 papers were identified for inclusion.

Notably, bidirectional encoder representations from transformers (BERT)-based studies were excluded based on three primary considerations: (1) the architecture of BERT is fundamentally designed for comprehension instead of generation tasks, which requires extensive fine-tuning (Devlin et al. (2018)). (2) This SLR emphasized models with inherent few-/zero-shot learning and advanced text generation capabilities. (3) This work focused on contemporary (2019–2025) developments in RE research that utilized LLMs (BERT (2018) represents an earlier generation of pretrained models).

- Inclusion criteria:
  1. Papers that focus on GenAI for RE
  2. Papers that are fully written in English
- Exclusion criteria:
  1. Papers that focus on GenAI for SE but not specifically on RE
  2. Papers that consider BERT as GenAI or LLM
  3. Gray and secondary literature such as book chapters, PhD theses, white papers, survey papers, and blogs

### 4.1.4 | Snowballing Process

To ensure the comprehensive coverage of relevant literature, we conducted backward and forward snowballing on the 238 papers selected using database searches. Backward snowballing involved examining the reference lists of the selected papers to identify additional relevant studies that might have been missed during database search. This process resulted in the identification and inclusion of four additional papers meeting the inclusion criteria.

Forward snowballing was performed to identify papers that cited the selected studies but did not yield additional papers for inclusion, which was attributed to several factors. (1) Our comprehensive multi-database search strategy had already captured the vast majority of relevant papers in this domain. (2) The relatively recent nature of research on GenAI use in RE means that many papers have not yet been extensively cited. (3) The search was conducted in May 2025, which limited the time available for forward citations to accumulate.

The final corpus, including the papers identified through snowballing, comprised 238 publications.

### 4.1.5 | Data Extraction

Data extraction was systematically conducted using standardized forms to ensure consistency across the selected papers. The extracted data included bibliographic information, research objectives, methodologies, and key findings, which were initially compiled in a Microsoft Excel spreadsheet. Data were subsequently migrated to Google Sheets to facilitate collaborative validation among the coauthors. This cloud-based approach enabled real-time review and quality control throughout the extraction process. The components of the data extraction framework consists of the categories and attributes, as summarized in Table 1, and the complete dataset is available in the Supplementary Materials.[‡]

[‡] https://github.com/haowei614/GenAI4RE_SLR_Data

**TABLE 1** Components of the data extraction framework

| Category | Attributes |
|---|---|
| Publication metadata | Authors; Title of publication; Year of publication; Publication venue; Phase(s) of RE addressed; Application domain; Task specificity. |
| GenAI model and configuration | Specific model and version; Implementation of fine-tuning protocols; Model configurations (parameters). |
| LLM application strategy and prompt engineering | LLM integration strategy; LLM application techniques; Learning paradigm; Prompt type; Prompt availability. |
| Outputs and resources | Tool or prototype name; Tool availability; Dataset or benchmark name; Dataset availability. |
| Evaluation | Evaluation metrics; Evaluation methodology; System quality attributions. |
| Literature quality assessment | General assessment of study quality and methodological rigor. |
| Challenges and future directions | Model-specific challenges (e.g., bias and hallucination); Industrial adoption status; Limitations. |

## 4.2 | Analysis and Classification Framework

### 4.2.1 | Publication Metadata

To facilitate the comprehension of the reviewed literature, we developed a systematic categorization framework that begins with the publication metadata. These metadata elements provide essential contextual information on the research landscape in GenAI for RE, which can enhance in understanding the temporal evolution, research distribution, and primary focus areas of the field. The following key factors were identified and scrutinized.

- **Author:** We documented the primary contributors to each article, which enabled the identification of influential researchers and research groups in the field.
- **Title:** The complete title of each publication was recorded to provide a clear overview of the research focus and facilitate literature tracking and citation.
- **Year:** The publication year was recorded to map the temporal evolution of GenAI applications in RE and to identify emerging trends.
- **Paper Type:** Studies were classified as full or short papers based on their respective conference submission guidelines.
- **RE Phase:** We categorized each study according to its primary focus within the RE lifecycle: requirements elicitation, analysis, specification, validation, and management. This categorization is based on the widely recognized RE process model by Pohl (Pohl (2010a)) in addition to an explicit analysis phase to emphasize the systematic processing of requirements between elicitation and specification.

  We categorized each study according to its primary focus within the RE lifecycle: requirements elicitation, analysis, specification, validation, and management. Our classification approach follows two rigorous methodological foundations: (1) Pohl's widely recognized RE process model, which provides the theoretical framework for the classification of RE activities (Pohl (2010a)) and (2) QA principles for SLR in SE (Yang et al. (2021)).
- **Application domain:** The domain in which the GenAI technique was applied (e.g., healthcare, finance).
- **Task Specificity:** Level of task specialization (generic RE support vs. task-specific applications).

### 4.2.2 | GenAI Model & Configuration

The technical aspects of GenAI applications were analyzed based on the specific LLMs used, fine-tuning strategies, and parameter configurations (especially temperature and prompt length) to reveal correlations between model configurations and performance in RE tasks.

- **Specific model and version:** We documented the specific GenAI architectures (e.g., GPT-3 and GPT-4) used in each study to track their adoption and effectiveness across the RE tasks.
- **Implementation of fine-tuning protocols:** We examined whether and how the models were fine-tuned for specific RE tasks and distinguished between studies that employ fine-tuning and directly use pretrained models (Howard and Ruder (2018)).
- **Model configurations (parameters):** We documented configuration details, such as temperature settings, and other relevant parameters that impact model performance in RE tasks (Holtzman et al. (2020)).

### 4.2.3 | LLM Application Strategy and Prompt Engineering

We extended the original prompt engineering analysis to include LLM integration strategies (e.g., hybrid or retrieval-augmented approaches) and application techniques. Prompt engineering was further classified according to learning paradigm, prompt type, and prompt availability to evaluate reproducibility and transparency in research practices.

- **LLM integration strategy:** Standalone use, hybrid integration (e.g., retrieval-augmented generation (RAG)-based approaches), or multi-model systems.
- **LLM application techniques:** Specific methods employed, such as retrieval-augmented generation or domain-specific embeddings.

**TABLE 2** Comprehensive QA results

| Assessment dimension | Mean | 1.0 | 0.5 | 0.0 |
|---|---|---|---|---|
| QA1: Research goal clarity | 0.996 | 237 (99.6%) | 1 (0.4%) | 0 (0%) |
| QA2: Research methodology appropriateness | 0.925 | 218 (91.6%) | 19 (8.0%) | 1 (0.4%) |
| QA3: Data analysis comprehensiveness | 0.842 | 194 (81.5%) | 40 (16.8%) | 4 (1.7%) |
| QA4: Adequacy of limitation discussion | 0.879 | 210 (88.2%) | 27 (11.3%) | 1 (0.4%) |
| **Overall score distribution (mean: 3.76/4.0)** | | | | |
| Score level | | Papers | | Quality pattern |
| High (4.0) | | 173 (72.7%) | | All indicators = 1.0 |
| Medium (3.5) | | 40 (16.8%) | | One indicator = 0.5 |
| Moderate (3.0) | | 20 (8.4%) | | Two indicators = 0.5 |
| Low ($\leq$ 2.5) | | 5 (2.1%) | | At least one indicator = 0 |

- **Learning paradigm:** We classified the approaches into zero-shot (no task-specific examples), one-shot (single example), few-shot (multiple limited examples), and CoT methodologies (Wei et al. (2022)). Figure 9 depicts the distribution of these paradigms across studies, highlighting the prevalent strategies for RE tasks under different data constraints.
- **Prompt type:** Four structural variations were identified, namely instruction-based (explicit directives), question-based (requirements as queries), description-based (contextual information), and iterative (progressive refinement) prompts. Figure 10 illustrates their distribution across the literature.
- **Prompt availability:** We assessed whether studies provided the exact text of the prompts used, which is essential for reproducibility (Filatovas et al. (2024)). This assessment helped identify gaps in research practices and promoted transparency in prompt engineering research.

### 4.2.4 | Outputs and Resources

The tools, prototypes, and datasets were systematically cataloged. Tool and dataset availability were analyzed to assess replicability and potential for industrial adoption.

- **Tool or prototype name:** Name of the developed tool or prototype, if any.
- **Tool availability:** Open-source, partially available, or proprietary.
- **Dataset or benchmark name:** Datasets used for training or evaluation.
- **Dataset availability:** Publicly accessible, partially available, or private.

### 4.2.5 | Evaluation

The evaluation approaches were classified according to metrics, methodologies, and quality characteristics as defined in ISO/IEC 25059, which facilitated a comparative analysis of how different studies assessed functional suitability, reliability, usability, and other quality attributes.

- **Evaluation metrics:** Metrics used to assess performance, such as accuracy, precision, recall, F1 score, or user satisfaction.
- **Evaluation methodology:** Experimental design, benchmark testing, user studies, or industrial case studies.
- **System quality attributions:** The ISO/IEC 25059 quality model was adopted as the analytical framework, including functional suitability, performance efficiency, compatibility, usability, reliability, security, maintainability, and portability (Cha (2019)). This standardized approach enabled a systematic comparison across GenAI techniques and provided insights into how different GenAI approaches enhance specific quality attributes of RE.

### 4.2.6 | Literature Quality Assessment

To ensure the reliability and methodological rigor of the reviewed studies, we conducted a structured QA following established guidelines in SE SLRs (Kitchenham et al. (7-15); Dybå and Dingsøyr (Dybå and Dingsøyr (2008))). The QA process evaluated four key criteria:

1. QA1: Does the paper clearly state the research goal and context?
2. QA2: Is the research methodology appropriate and rigorously applied to meet the stated objectives?
3. QA3: Is the data analysis comprehensive and are the results thoroughly presented?
4. QA4: Are the limitations adequately discussed and meaningful suggestions provided for future research?

Each study was scored on a three-point scale (yes = 1.0, partial = 0.5, no = 0), which resulted in a total score range of 0–4. Three researchers independently assessed each paper, and discrepancies were resolved through consensus to ensure scoring reliability.

The results of the assessment are summarized in Figure 4 and Table 2. Overall, the literature exhibited high methodological quality:

1. QA1 (goal clarity) achieved a mean score of 0.998, with nearly all papers (99.6%) scoring 1.0.
2. QA2 (methodological appropriateness) had a mean of 0.956, confirming that most studies applied systematic methods.

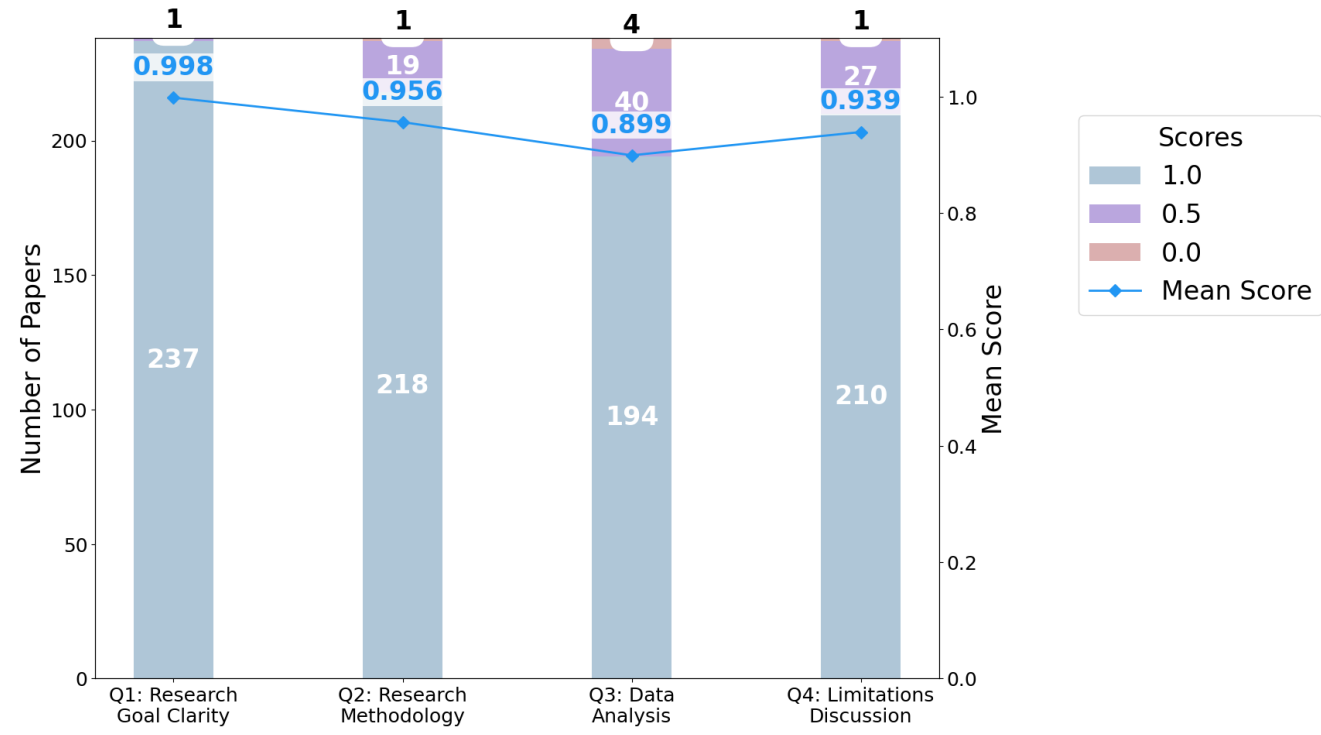

**FIGURE 4** Overall score distribution.

3. QA3 (data analysis) and QA4 (limitation discussion) showed slightly lower mean scores (0.899 and 0.939 respectively), indicating opportunities for deeper analysis and more explicit reflections on study boundaries.

The overall QA mean score across 238 papers was 3.79/4.0, with 72.7% of papers achieving the maximum score. Only 2.1% of papers scored below 2.5.

These results demonstrate a strong baseline of academic rigor in GenAI–RE literature. However, the relative weakness in QA3 and QA4 suggests the need for improved data reporting transparency and a more critical engagement with study limitations.

This QA evaluation provided the foundation for our synthesis in Section 5 and informed the upcoming classification of methodological and systemic challenges (Section 4.2.7). Specifically, the variability in analytical depth and underreporting of validity threats motivated us to emphasize reproducibility, evaluation design, and governance structures in the roadmap (Section 7).

### 4.2.7 | Challenges and Future Directions

- **Model-specific challenges:** To systematically analyze the challenges and limitations of the application of GenAI in RE, we developed an evaluation framework based on 10 critical aspects commonly discussed in LLM literature. This framework was adapted from the comprehensive survey conducted by Naveed et al. (Naveed et al. (2024)), which we specifically tailored for the RE context. The aspects examined include the following:
    - Bias and fairness in model outputs;
    - Ethical and regulatory considerations;
    - Security and privacy concerns;
    - Model interpretability and explainability;
    - Computational and economic costs;
    - Real-time processing capabilities;
    - Hallucination prevention and mitigation;
    - Result reproducibility;
    - Model controllability; and
    - Authorship and copyright issues.
- **Industrial adoption status:** To evaluate the real-world maturity of GenAI applications in RE, we analyzed the industrial adoption status of each reviewed study. Rather than relying on a predefined taxonomy, we employed an inductive approach in which classifications were derived directly from the reported implementation details and deployment contexts observed across the literature corpus.
- **Limitations:** In addition to evaluating model-level challenges, we analyzed the explicit limitations reported by the reviewed studies. These limitations reflected the authors' own recognition of methodological, contextual, or operational constraints and provided insights into the recurring obstacles hindering the use of GenAI in RE.

## 5 | RESULTS, ANALYSIS, AND DISCUSSION

Building on the analysis framework discussed in Section 4, this section presents the SLR findings organized around the RQs. We examined the application of GenAI technologies across different RE phases, analyzed research trends and methodologies employed, assessed the quality of the literature, and identified challenges and opportunities in the field. The analysis provided comprehensive insights into how GenAI is shaping contemporary RE practices.

### 5.1 | RQ1: Publication Trends pertaining to GenAI for RE

Based on the respective conference/journal venue standards, the screened papers were classified into 230 full papers and 8 short papers. We also categorized the papers according to their focus within the RE lifecycle.

As shown in Figure 5, requirement analysis received the highest attention (30.0%), followed by elicitation (22.1%) and specification (22.1%). Studies on requirement validation had a share of 19.0%, and requirement management remained underexplored (6.8%).

This distribution suggests that current research predominantly emphasize analytical tasks, possibly because GenAI techniques (e.g., LLMs for text mining and clustering) are well suited for processing and analyzing large volumes of requirement-related data. Similarly, elicitation and specification benefit from the natural language understanding and generation capabilities of GenAI. In contrast, the limited focus on management may reflect the lack of mature AI-driven tools for project-wide requirement tracking and evolution, which often involve complex sociotechnical factors rather than purely linguistic ones.

The temporal distribution of publications (Figure 6) indicates that the use of GenAI in RE has experienced exponential growth. The publication volume increased from four papers in 2022 to 23 papers in 2023, further increasing to 113 papers in 2024. Although the 2025 data represents partial year coverage (up to May), the current count of 97 papers suggests that the annual total will likely surpass that of 2024.

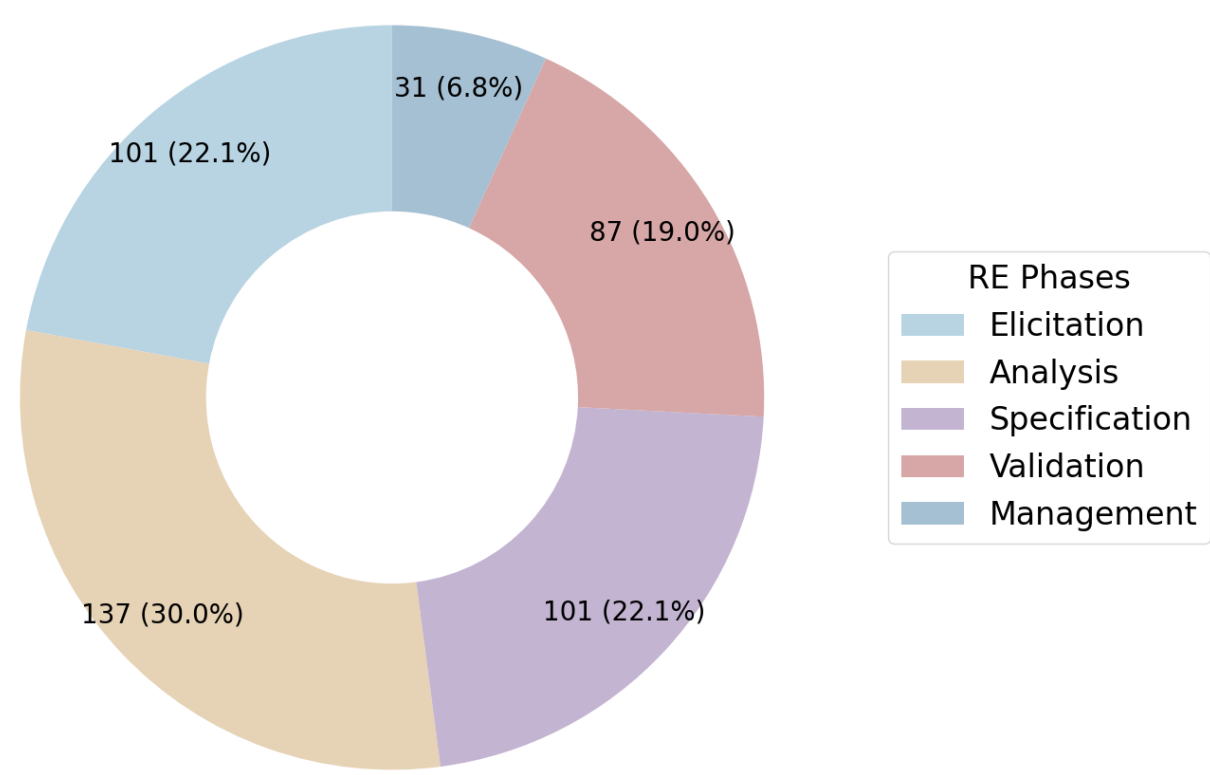

**FIGURE 5** Distribution of RE phases (N=238).

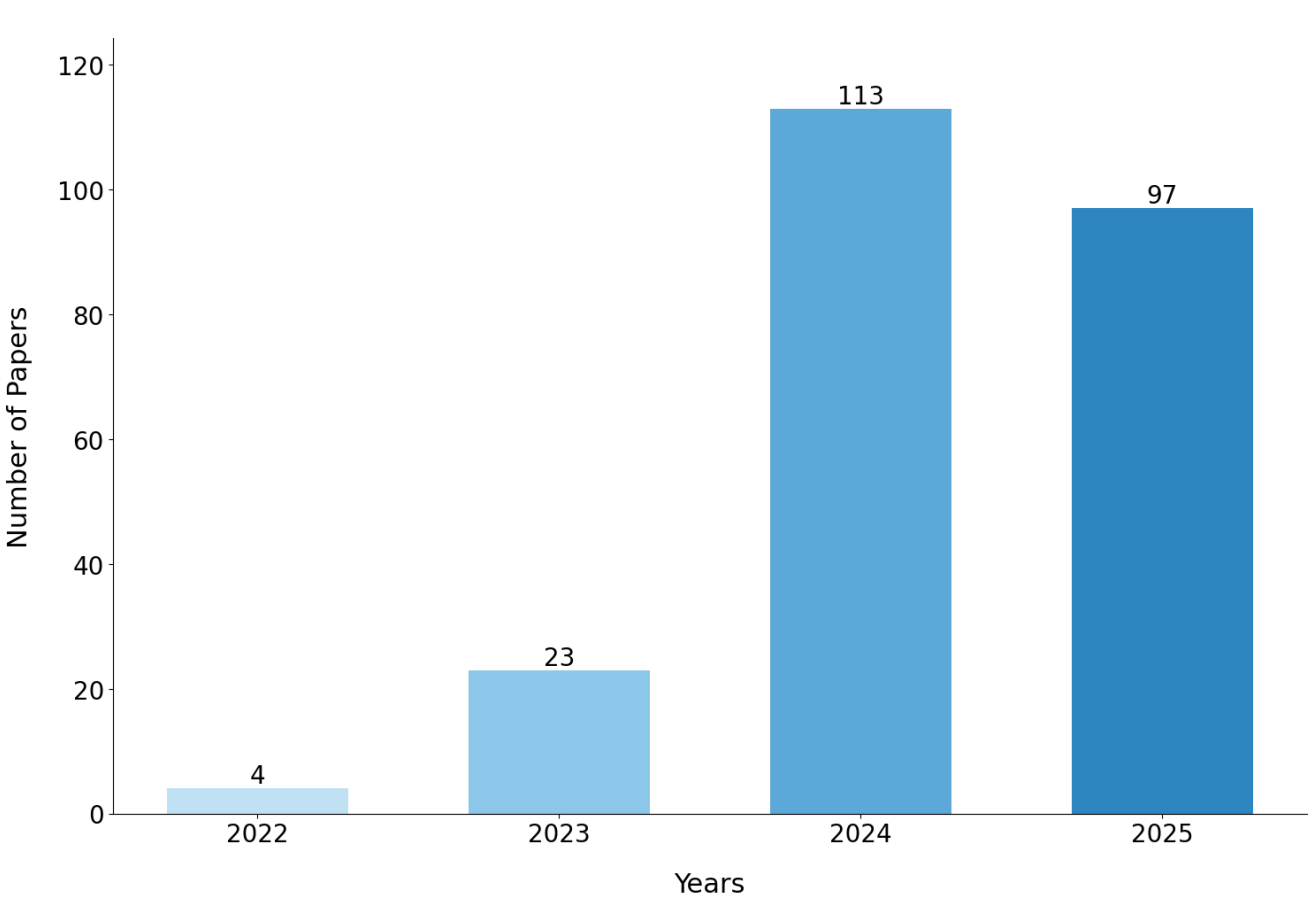

**FIGURE 6** Distribution of papers across years (N=238).

This growth trajectory reflects the mainstream adoption of LLMs following breakthrough developments such as the release of ChatGPT in late 2022, increasing recognition of GenAI potential in RE practices, and maturation of foundational AI technologies. The steep acceleration from 2022 to 2024 further indicates that the use of GenAI in RE has transitioned from an emerging niche to a rapidly consolidating research domain.

**Main Findings for RQ1**

The analysis highlights several noteworthy trends in the use of GenAI in RE. First, this field has undergone rapid expansion, transitioning from an emerging niche to a mainstream research area. Since 2023, the number of publications has increased exponentially, which reflects the growing interest in leveraging GenAI to address long-standing RE challenges and indicates a shift from purely exploratory studies to more specialized and methodologically rigorous investigations.

Second, the research focus is unevenly distributed across the RE phases. Most studies concentrate on requirements analysis, elicitation, and specification, which aligns with the NLP and generation capabilities of GenAI. In contrast, requirements management and validation are underexplored, which suggests that GenAI has yet to be fully adapted to the complex socio technical and process-oriented aspects of RE. Furthermore, the majority of existing studies are academic and experimental in nature, with empirical evidence from industrial contexts remaining limited.

**Takeaway Message 1**

**For researchers:** Future studies should address the imbalance in research across RE phases, particularly the underrepresentation of requirements management and validation. Beyond text-centric applications, promising directions include integrating GenAI with sociotechnical and process-oriented RE practices (e.g., change impact analysis, automated traceability). Moreover, as the field matures, rigorous empirical evaluations, benchmarking, and domain-specific adaptations will become essential for enabling reliable industrial adoption.

**For practitioners:** The concentration of research on analysis, elicitation, and specification indicates that GenAI tools for these phases are becoming increasingly robust and can be piloted in industrial settings, especially for automated document generation and requirement classification. However, for management and validation, GenAI should currently be used as an assistive rather than autonomous tool, given the lack of mature solutions. The exponential growth of the number of publications underscores the need for the continuous monitoring of new developments to identify validated and practically deployable solutions.

## 5.2 | RQ2: Methodology Trends

### 5.2.1 | Comprehensive Analysis of GenAI Models and Techniques

Initially, we observed concentration in model selection patterns. As shown in Figure7, GPT-based models dominating the landscape. The GPT-4 series accounted for 36.7% of the implementations and was primarily applied in complex requirement analysis (78% of GPT-4 studies) and multistakeholder elicitation scenarios, where its superior contextual understanding justified the 3-5 times higher operational costs compared with those of smaller models. The GPT-4 series was followed by the GPT-3.5 series (25.3%), which demonstrated a comparable performance in routine classification tasks while offering better cost-effectiveness for large-scale deployment. This finding indicates a strong preference for high-capacity proprietary models. Other proprietary models, such as Claude, Gemini, and PaLM, had a share of 20.7%, while open-source alternatives, including LLaMA and CodeLlama, represented only 11.6%. Despite its low adoption percentage, CodeLlama showed specialized effectiveness in code-requirement traceability tasks, with eight studies achieving 85% accuracy in automated traceability matrix generation and 23% lower hallucination rates in technical specifications compared with general-purpose models. This distribution suggests potential overreliance on a single family of models, which may limit the exploration of diverse approaches to address RE-specific challenges. However, our analysis reveals distinct performance characteristics, with GPT models excelling in linguistic complexity and specialized models such as CodeLlama providing superior technical accuracy in code-related RE tasks.

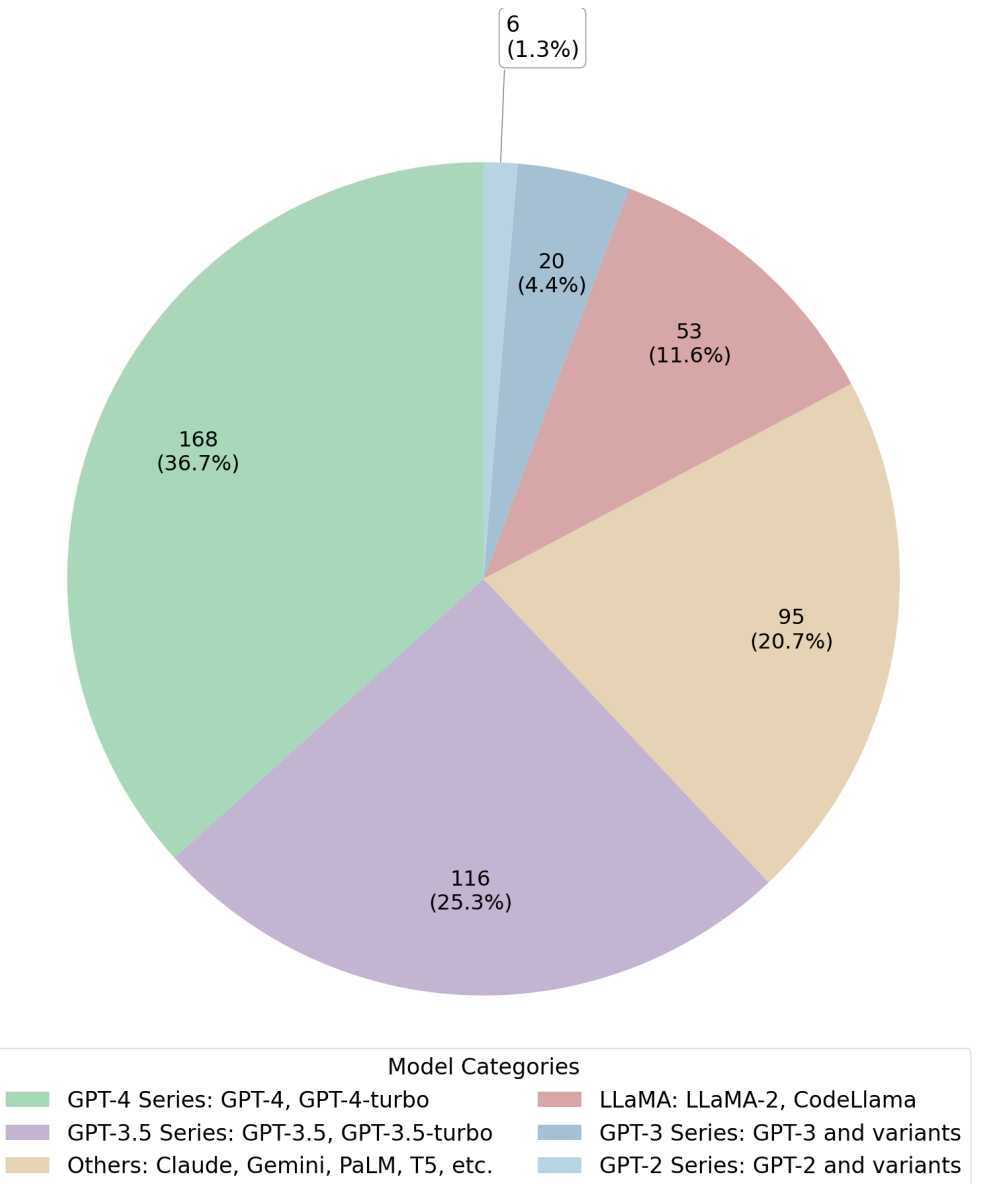

**FIGURE 7** Distribution of GenAI models (N=238).

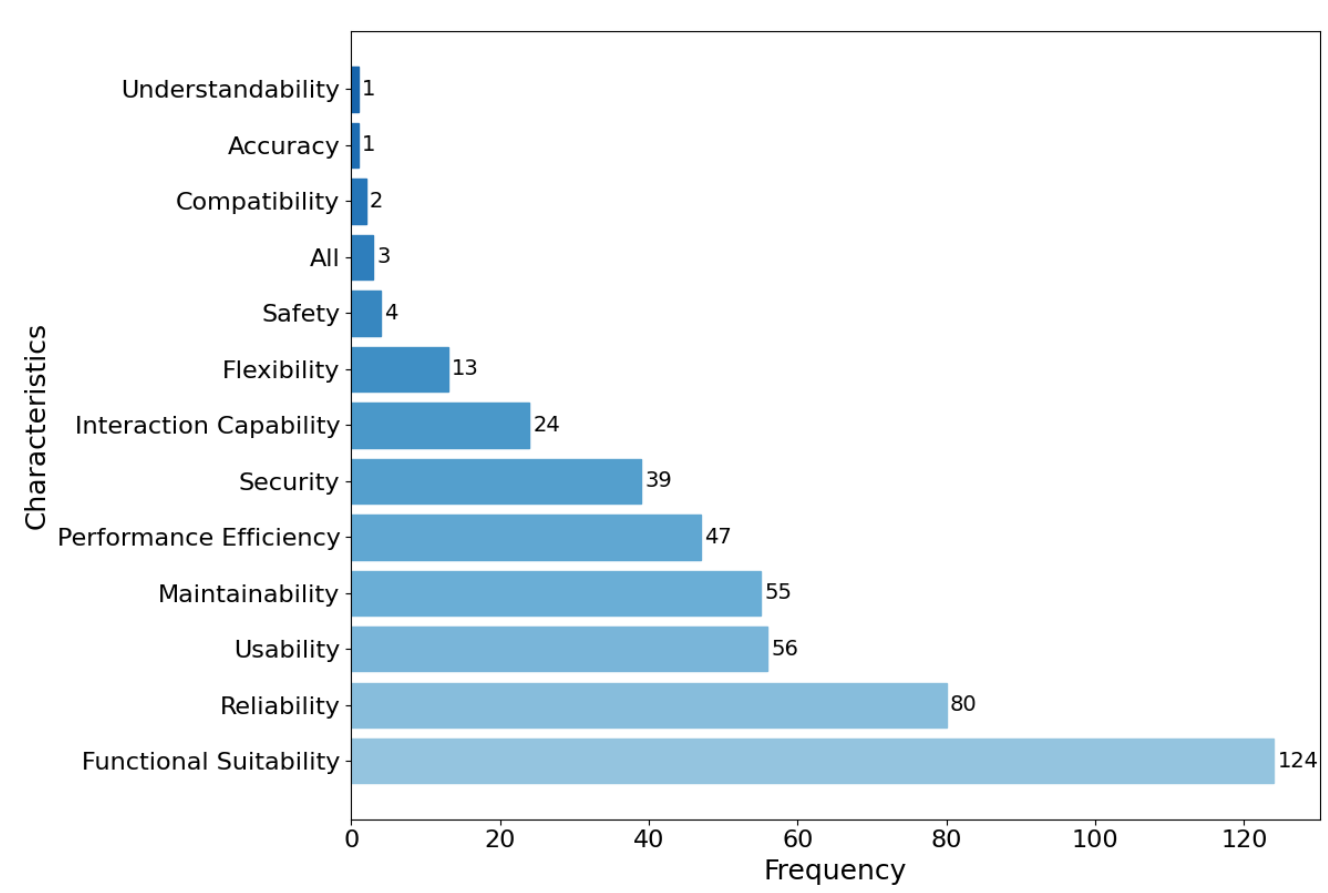

**FIGURE 8** Distribution of software quality characteristics.

The analysis of software quality characteristics highlighted the prioritization of operational performance (Figure 8). Functional suitability dominated (124 occurrences) and was followed by reliability (80 occurrences), usability (56 occurrences), and maintainability (55 occurrences). In contrast, long-term and system-level qualities received limited attention: Security was mentioned 39 times, interaction capability 24 times, and flexibility only 13 times, and aspects such as accuracy and understandability were almost neglected. This imbalance indicates that current research primarily emphasizes immediate functional effectiveness rather than long-term sustainability, robustness, or user-centered qualities.

Finally, implementation approaches demonstrated a high preference for pretrained models, with >90% of the studies employing models without domain-specific fine-tuning. This trend is seemingly driven by several factors, including the effectiveness of transfer learning in RE tasks, resource constraints in research settings, and limited availability of domain-specific training data. Although the use of pretrained models enables rapid deployment and experimentation, it may constrain the potential for specialized optimization in domain-specific contexts. The analysis of parameter configuration practices revealed a consistent pattern favoring conservative settings. Studies have focused on temperature parameters, featuring a preference for low values (0–0.4) to prioritize output consistency (Liu et al. (2024)). This narrow focus on a single parameter, combined with the limited documentation of other configuration options, indicates a potential underexploration of model capabilities. The prevalence of default configurations for other parameters points to an opportunity for a more systematic investigation of parameter optimization strategies.

### 5.2.2 | Prompt Engineering

The comprehensive analysis of prompt engineering methodologies across the reviewed papers revealed patterns and emerging research trends in leveraging GenAI for RE tasks. The analysis framework encompassed three critical dimensions, namely learning paradigms, prompt formats, and prompt availability.

As shown in Figure 9, the distribution of learning paradigms adopted for the RE tasks demonstrated a preference for approaches balancing model guidance with generalization capabilities. The dominance of few-shot learning (43.6% of studies) confirmed its effectiveness in improving model performance through limited but targeted examples, particularly for complex requirement analysis tasks. Comparative analysis revealed that few-shot approaches consistently outperformed zero-shot methods by 15%-25% in requirement classification accuracy, with studies by Hassani et al. (2024) and Görgen et al. (2024), demonstrating superior performance in domain-specific contexts. Zero-shot learning followed at 37.7%, indicating success in direct task execution without prior examples, which suits relatively straightforward RE tasks despite the notably lower precision in complex analytical scenarios. CoT approaches has a share of 14.0%, which reflected the growing but still limited adoption of explicit reasoning steps to enhance interpretability in complex reasoning scenarios. Our analysis showed that CoT prompting delivers marked improvements in requirement conflict detection and identified inconsistencies better than standard prompting approaches. One-shot learning has a share of 4.7%, which highlights its limited practical utility in RE contexts due to insufficient contextual guidance for complex reasoning tasks. This distribution suggests that researchers tend to rely on the inherent reasoning capabilities of large models (zero-shot) or provide multiple examples for targeted guidance (few-shot), with intermediate solutions such as one-shot receiving little attention.

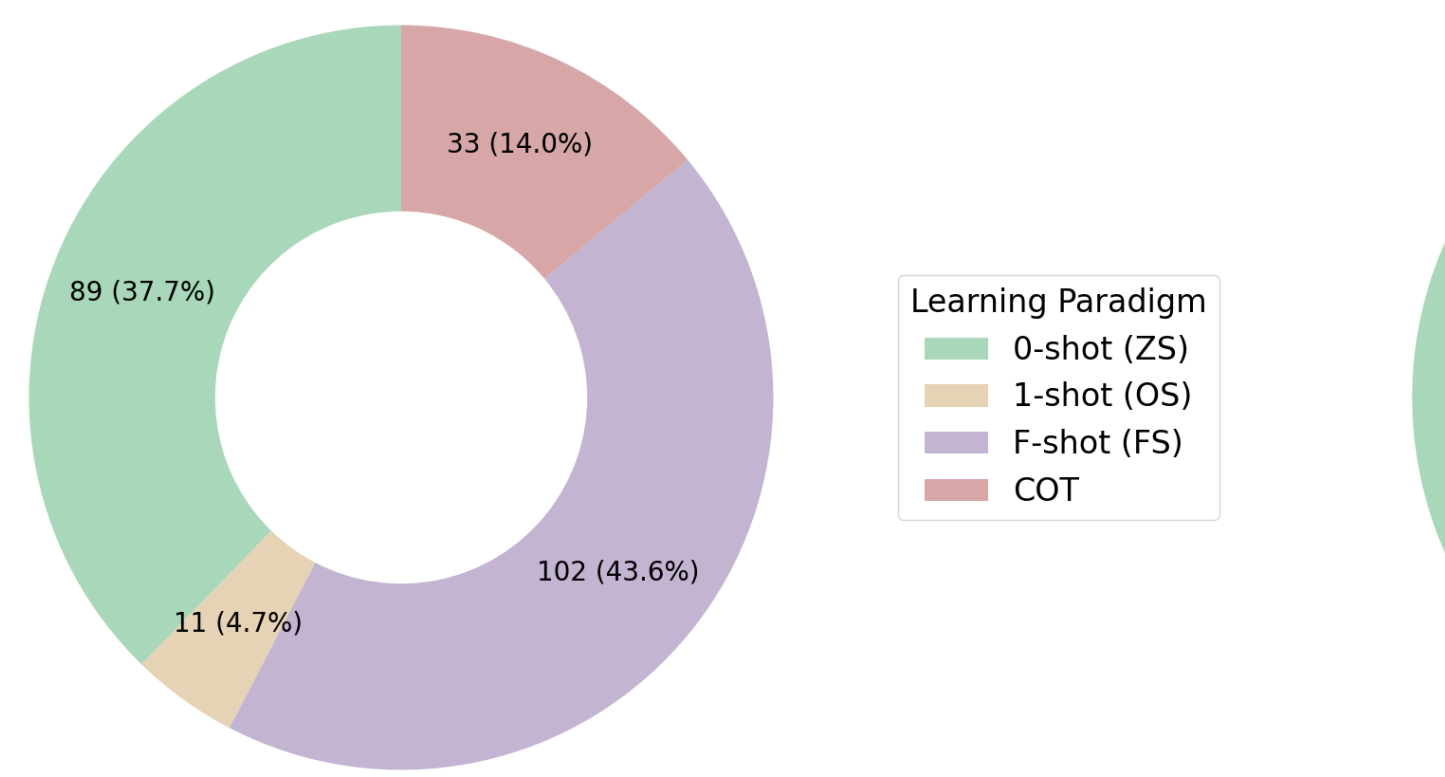

**FIGURE 9** Distribution of learning paradigm (N=238)

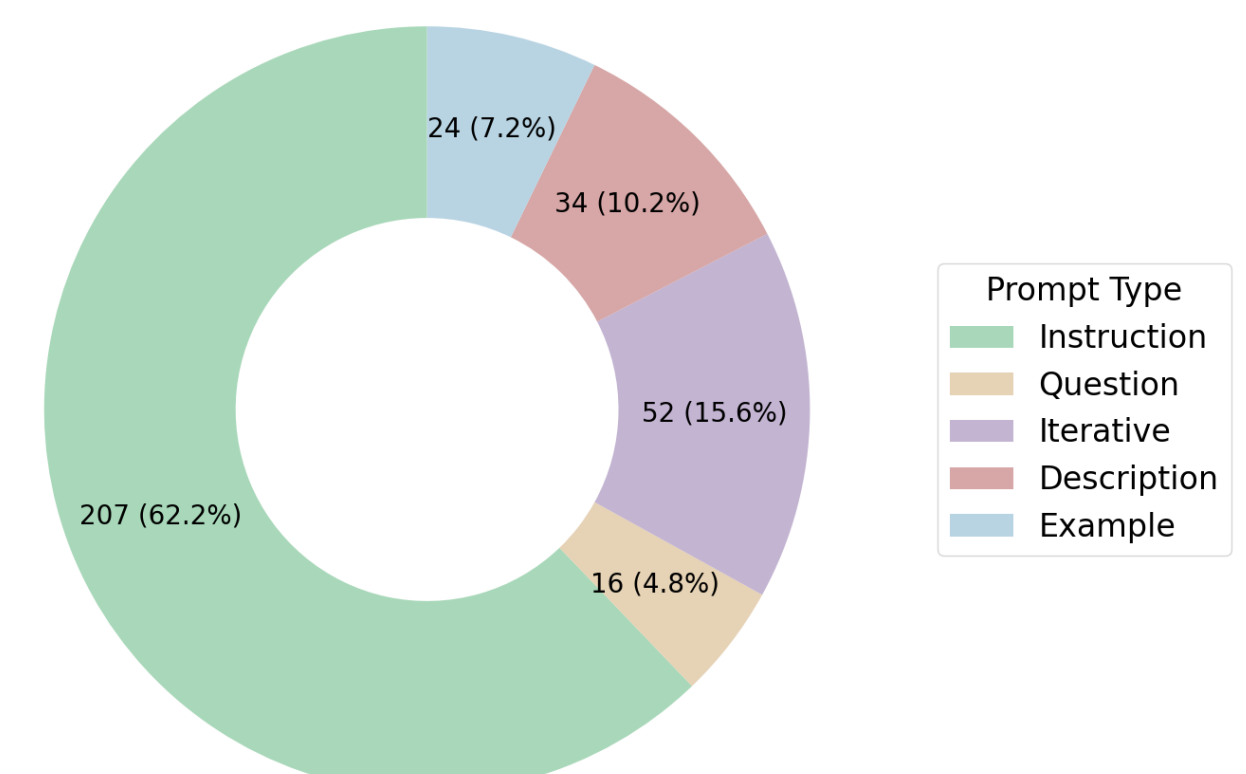

**FIGURE 10** Distribution of prompt types (N=238).

The dominance of instruction-based prompts (62.2% of occurrences) reflected the highly structured nature of RE tasks and effectiveness of providing direct explicit guidance to GenAI models (Figure 10). Iterative prompting followed (15.6%), emphasizing the importance of progressive refinement and feedback loops for generating accurate and contextually appropriate requirements. Question-based formats accounted for 10.2%, mainly supporting requirement elicitation through interactive querying, while example-based prompts (7.2%) were found to be typically employed as supplementary cues to improve the ability of the model to understand task-specific contexts. Description-based approaches were the least adopted (4.8%), which indicated the limited use of narrative-style contextualization in RE tasks. This distribution suggests a reliance on well-structured and directive prompting strategies, which are perceived as more reliable for ensuring consistent outputs in RE. However, the low adoption of interactive (question-based) and example-based strategies highlights an underexplored opportunity for developing more adaptive and user-centric prompting methods. Furthermore, the minimal use of description-based prompts suggests that current research has yet to fully exploit contextual and semantic-rich representations, which could enhance the interpretability and reasoning performance of GenAI in complex RE scenarios.

Prompt availability analysis revealed a trend toward methodological transparency. Among the reviewed studies, >80% explicitly provided prompt examples, whereas 20% did not disclose their prompts. This high level of availability suggests that the field is moving toward increased reproducibility and standardization, which facilitates result validation and comparison across studies.

However, the remaining proportion of unavailable prompts highlights persistent challenges, such as proprietary restrictions, data privacy concerns, or the use of ad hoc experimental setups that are not easily generalizable. This indicates the need for community-driven guidelines on prompt sharing, which can further enhance reproducibility and foster the development of standardized benchmarking practices in the use of GenAI in RE.

### 5.2.3 | Model Enhancement Techniques

While GPT series models dominate the field with an adoption rate of 67.3%, researchers have explored various enhancement techniques to address the specific challenges of RE tasks. Analysis of implementation patterns revealed that 11 papers (e.g., Lian et al. (2024), Cheng et al. (2024), and Wei (2024)) employed fine-tuning techniques to adapt models to domain-specific requirements tasks. Additionally, 30 papers (e.g., Alawaji et al. (2024), Alharbi et al. (2024), and Han et al. (2024)) explored multimodel integration approaches, combining different strengths of the models to improve performance in specialized RE contexts.

Of particular interest is the emerging use of RAG, as demonstrated in (Arora et al. (2024)) in which the researchers developed RAG-based test scenario generation (RAGTAG), a tool for the generation of automated test scenarios from natural language requirements. Their industrial evaluation at the Austrian Post demonstrated that RAG integration improved the accuracy and reliability of generated test scenarios by grounding them in domain-specific knowledge. The tool was particularly effective in handling bilingual requirements (German and English) and demonstrated a strong performance in test scenario coverage and feasibility. This approach represents an advancement in addressing hallucination challenges and enhancing the reliability of generated requirements.

Our analysis also revealed innovative approaches to external knowledge integration, in which studies such as Hou et al. (2024), Sainio et al. (2024), and Belzner et al. (2024) demonstrate methods of incorporating domain-specific knowledge and specialized RE tools. For instance, Hou et al. (2024) utilized the m3e-base model for embedding requirements information, while Sainio et al. (2024) developed specialized patterns for requirement management integration. Other studies opted for a hybrid approach, which combines multiple enhancement techniques. For example, Hassani et al. (2024) not only applied fine-tuning but also integrated multiple models (Phi-2, Mistral, Zephyr, GPT-3.5 Turbo, and GPT-4) to leverage their complementary strengths. This pattern of technical enhancement reveals opportunities and challenges in the field. While basic model applications are common, sophisticated enhancement approaches, such as RAG and fine-tuning, remain underutilized despite their potential benefits. The selective adoption of enhancement techniques is seemingly influenced by factors such as computational requirements, data availability, and implementation complexity. This finding suggests an opportunity for future research to develop more accessible and efficient enhancement techniques specifically tailored to RE tasks, particularly in areas in which traditional LLM approaches face limitations. The success of studies that implemented these enhancements indicates their value in improving model performance, especially in critical RE tasks that require high levels of accuracy and reliability.

### 5.2.4 | LLM Application Strategy

The analysis of LLM application strategies revealed a preference for standalone approaches. Most studies were found to rely on LLMs as independent tools without additional external components, with hybrid strategies integrating RAG, external tools, or multiagent frameworks remaining uncommon. This finding indicates that current research largely focuses on validating the inherent capabilities of LLMs rather than exploring more complex integrated architectures.

Similarly, the investigation of LLM application techniques revealed the dominance of zero-shot prompting, which reflected a strong reliance on the pretrained knowledge of large models for RE tasks. Few-shot prompting appeared less frequently, and advanced reasoning techniques such as CoT, tree-of-thought (ToT), or reinforcement learning from human feedback (RLHF) were rarely adopted. Human-in-the-loop and role-based prompting methods are also underexplored.

These patterns suggest that the use of GenAI-for-RE is still in an early methodological stage, with limited experimentation performed in advanced or hybridized approaches. Although the dominance of standalone and zero-shot strategies facilitates rapid deployment and experimentation, future research should explore hybrid architectures, retrieval-enhanced methods, and interactive prompting strategies to improve reasoning quality and adaptability in complex RE scenarios.

**Main Findings for RQ2**

First, model selection shows a strong preference for proprietary GPT-based models (>60%), with the exploration of alternative or open-source models remaining limited. This overreliance may constrain innovation and adaptability to RE-specific challenges.

Second, software quality considerations are skewed toward immediate operational performance, with functional suitability and reliability dominating, whereas long-term and system-level qualities—such as security, maintainability, and interpretability remain underexplored.

Third, implementation strategies favor pretrained models and conservative configurations, with >90% of studies avoiding domain-specific fine-tuning and advanced parameter optimization, which reflects an early-stage methodological maturity.

Fourth, prompt engineering is dominated by instruction-based formats (62.2%) and few-shot paradigms (43.6%), which reflects a preference for structured and example-guided prompting. Although >80% of studies share prompt details, interactive strategies (e.g., question-based prompts and human-in-the-loop methods) and reasoning-enhanced approaches (e.g., CoT and ToT) remain rare. Collectively, these patterns reveal the prioritization of rapid deployment and baseline functional performance, with the adoption of advanced techniques aimed at improving reasoning depth, reliability, and domain-specific adaptability remaining limited.

**Takeaway Message 2**

**For researchers:** The current technical landscape of GenAI use in RE reflects the tension between rapid adoption and methodological immaturity. The overwhelming reliance on proprietary GPT-based models and dominance of zero- and few-shot prompting indicate that research remains largely exploratory and primarily focuses on validating the generic capabilities of LLMs rather than designing RE-specific solutions. This overreliance is not only a technical convenience but also a symptom of data scarcity, high computational costs, and the lack of standardized evaluation protocols, which collectively discourage experimentation with alternative models, domain-specific fine-tuning, or parameter-efficient optimization methods such as PEFT and QLoRA. Similarly, the limited use of advanced reasoning techniques (CoT, ToT, RLHF) reveals a deeper methodological gap, i.e., current research tends to treat LLMs as "black-box text generators" rather than reasoning agents that can be systematically engineered for requirement analysis, conflict resolution, or traceability.

Quality assurance (QA) considerations further expose this immaturity. The overwhelming focus on functional suitability and reliability suggests that current evaluations are task-oriented and short-term, whereas long-term system-level qualities such as security, maintainability, transparency, and interpretability are almost absent from empirical investigations. Without these considerations, the integration of GenAI integration into safety-critical or regulated domains will remain limited. To move forward, researchers must shift from ad hoc task evaluations to a layered research agenda, i.e., (1) diversify technical strategies through hybrid architectures that embed RE-specific knowledge (e.g., RAG, knowledge graph integration, or multi-agent collaboration), (2) develop cost-efficient fine-tuning pipelines and adaptive prompting frameworks tailored to RE complexities, and (3) establish domain-specific QA benchmarks aligning with ISO/IEC 25059 and enabling rigorous comparative evaluations of GenAI solutions. The high prompt availability across studies provides an unprecedented opportunity to bootstrap such benchmarking, provided that community-driven standards for prompt documentation and sharing are established.

**For practitioners:** From a practical standpoint, current generation of GenAI tools are best viewed as assistive accelerators rather than autonomous decision-makers for RE. Their strength lies in well-structured low-risk tasks such as automated drafting, requirement classification, and basic consistency checking, where high functional suitability and reliability are sufficient. However, the lack of robust evidence for long-term qualities, particularly security, maintainability, and interpretability makes them unsuitable for mission-critical, safety-regulated, or compliance-heavy RE contexts without human oversight.

Nevertheless, the field is at a turning point. Early industrial experiments with RAG and hybrid approaches have shown that grounding LLM outputs in domain-specific knowledge can markedly improve reliability, reduce hallucinations, and even support multilingual requirement processing. This trend signals a shift from generic LLM use to context-aware domain-adapted GenAI solutions, which are likely to define the next wave of industrial adoption. Therefore, practitioners should adopt a dual strategy, i.e., leverage existing GenAI tools for low-risk automation while actively monitoring and participating in the development of hybrid knowledge-integrated solutions that promise higher reliability and regulatory compliance.

## 5.3 | RQ3: Evaluation Landscape

To understand how LLMs are evaluated in the context of RE, we analyzed four key dimensions, namely tool and dataset availability, the use of benchmarks, evaluation metrics, and methodological strategies, with the findings presented in Figures 11 and 12.

The landscape of tool and dataset availability reveals encouraging trends and persistent gaps (Figure 11). Among the reviewed studies, 57 (23.9%) explicitly released their tools, although a larger number (174; 73.1%) did not make tools publicly accessible or disclose their availability. Similarly, although 116 studies (48.7%) employed or constructed datasets for evaluation, only a subset ensured dataset availability. Notably, 109 studies (45.8%) used publicly unavailable datasets, and a small proportion indicated partial or unspecified availability. This asymmetry between usage and sharing indicates that although the field is shifting toward implementation-oriented research, it has not yet fully embraced the open-science practices required for reproducibility and comparative evaluation. These findings underscore a critical challenge for the GenAI-for-RE community, showing that the lack of shared tooling and benchmarking infrastructure impedes cumulative knowledge building and slows methodological maturation.

Figure 12 illustrates the distribution of evaluation metrics and methodologies, focusing on the five most prevalent categories becasue of the high variance across studies. In terms of metrics, precision/recall/F1 score were identified as the most frequently used metrics (119 studies), particularly in classification and extraction tasks. Accuracy-based measures appeared in 40 studies, while 22 studies employed human evaluation to assess aspects such as coherence or usability. However, more rigorous evaluation protocols, including error analysis (11 studies) and NLP-specific or semantic similarity metrics (fewer than 10 studies), were rarely adopted. The dominance of surface-level performance metrics reflects a concerning trend, revealing that although quantitative indicators are readily computable, they may fail to capture the complexity and domain specificity of RE tasks.

Regarding methodology, most studies adopted controlled experiments (120) or case studies (43), which reflects the grounding of the field in SE experimentation. Comparative studies (15), user studies (12), and empirical deployments (7) remained in the minority. The rare use of surveys or interviews (3) and near-absence of mixed-method triangulation highlight the lack of methodological pluralism. More importantly, few studies triangulated human and quantitative evaluations or reported interrater reliability, which limits external and construct validity. This trend suggests that the practices for evaluating the use of GenAI in RE remain ad hoc and task-specific, lacking the consistency, robustness, and transparency required for large-scale generalization or industrial certification.

Together, these patterns highlight a field in transition. On the one hand, the increasing availability of tools and datasets, combined with the diversity of evaluation approaches, points toward a more empirically grounded research culture. However, the lack of standardized benchmarks, inconsistent sharing practices, and minimal methodological triangulation reflects deeper systemic limitations. To support cumulative innovation and enable meaningful cross-study comparisons, future work must move toward a more coordinated infrastructure for evaluation that combines quantitative rigor, domain-relevant metrics, human-centered insight, and publicly available resources. Only through such integrative practices can the field advance toward trustworthy, scalable, and replicable GenAI solutions for RE.

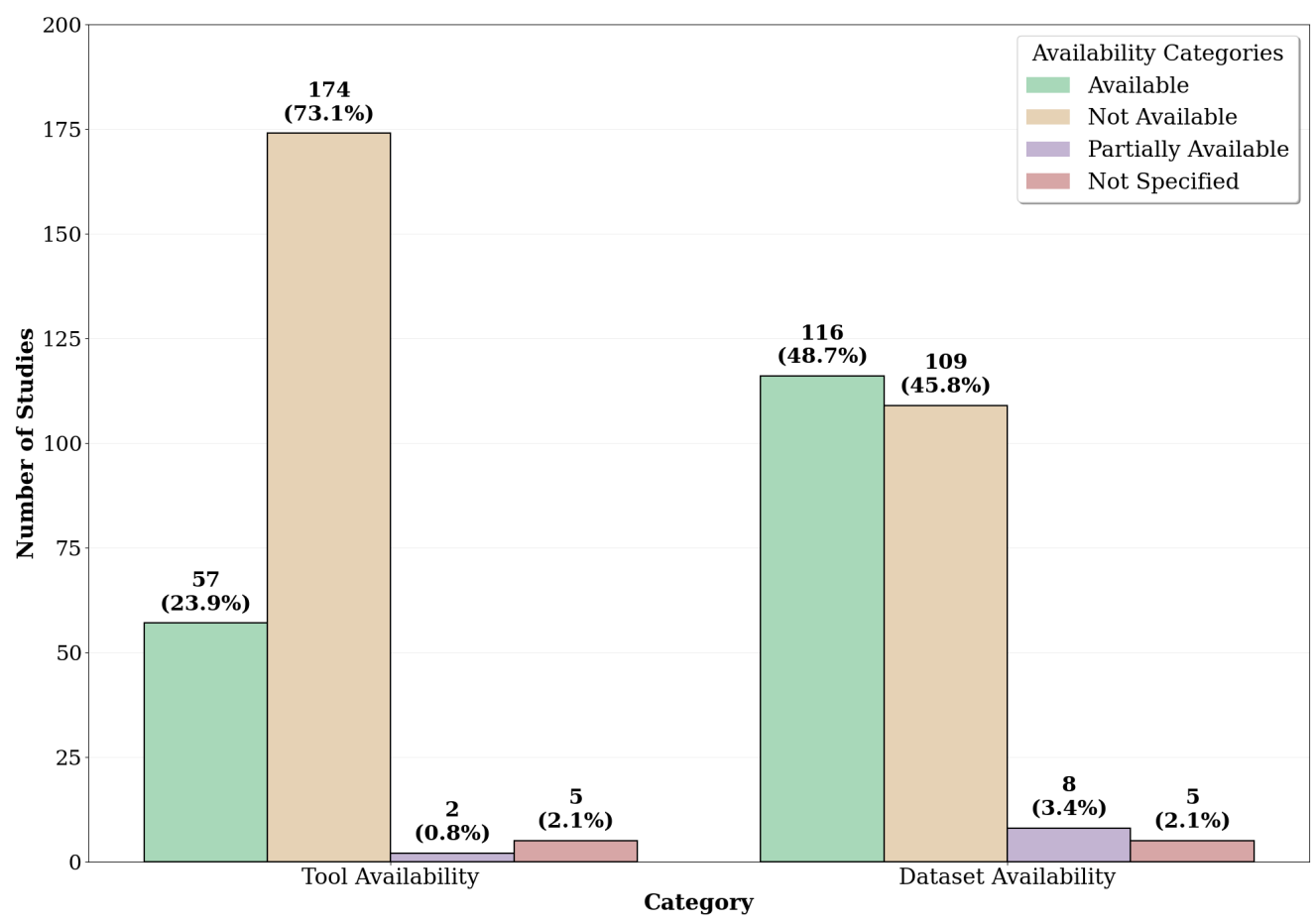

**FIGURE 11** Distribution of tool and dataset availability (N=238).

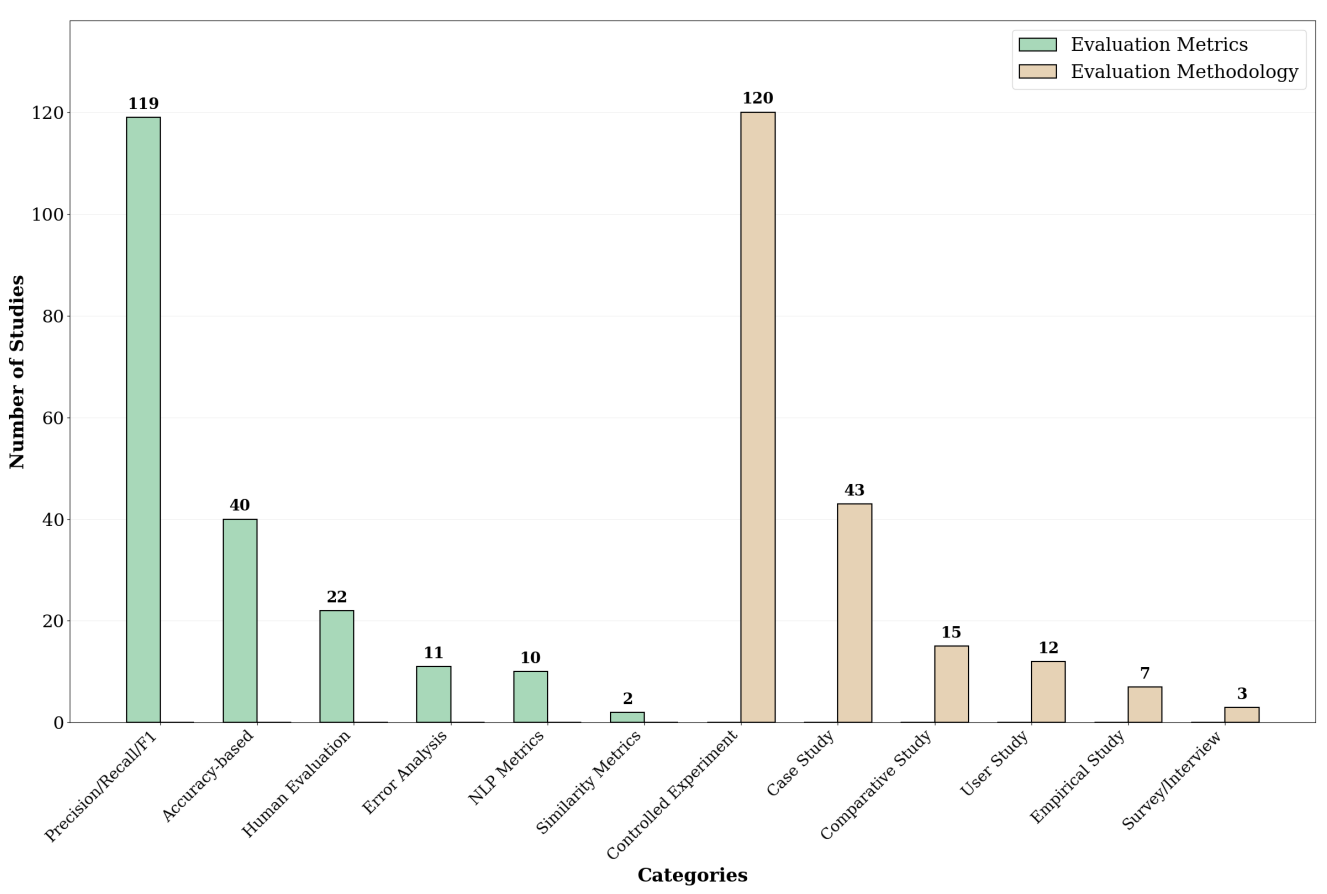

**FIGURE 12** Distribution of evaluation metrics and methodology (N=238).

### Main Findings for RQ3

First, although a growing number of studies have reported the development of tools and use of datasets, actual availability remains limited. Only a minority of the tools and datasets are openly shared, which reflects a disconnect between implementation-oriented research and open-science practices. This lack of transparency impedes reproducibility and comparative benchmarking and signals a structural weakness of the current research ecosystem.

Second, evaluation metrics remain concentrated around surface-level quantitative indicators, such as precision, recall, and F1 score, with the adoption of richer or domain-specific metrics remaining minimal. Human evaluation is employed in some studies but rarely triangulated with quantitative measures, and most studies do not report interrater reliability or statistical validation.

Third, the methodological strategies for evaluation are uneven and fragmented. Although controlled experiments and case studies are common, more robust or hybrid empirical designs such as mixed-method studies, longitudinal deployments, and structured user feedback protocols remain rare. This points to a methodological immaturity of the validation of GenAI systems for RE tasks and reveals a critical need for more rigorous, transparent, and human-centered evaluation frameworks.

### Takeaway Message 3

**For researchers:** The current evaluation practices in the use of GenAI-for-RE are limited by fragmentation, superficiality, and limited reproducibility. The absence of open-access tools and datasets not only limits research transparency but also stalls cumulative progress by preventing benchmarking and replication. Moreover, the narrow focus on conventional NLP metrics ignores the complex sociotechnical nature of RE, where interpretability, stakeholder alignment, and process integration are equally critical. To address these gaps, researchers must pivot toward a more principled evaluation paradigm that incorporates (1) domain-grounded evaluation criteria (e.g., traceability and ambiguity resolution), (2) mixed-method designs combining quantitative metrics with practitioner insights, and (3) community-standardized benchmarks with open data and prompt-sharing protocols. Such efforts will not only strengthen methodological validity but also bridge the gap between academic experimentation and industrial deployment.

**For practitioners:** Although the increasing availability of tools and RE datasets suggests a transition to practical application, practitioners should remain cautious in interpreting the reported results. The dominance of surface-level metrics and one-off experiments indicates that many current studies lack the depth necessary for deployment in high-stake, regulatory-bound RE contexts. Therefore, practitioners should demand greater methodological transparency, including access to prompts, evaluation datasets, and human judgment protocols, when assessing GenAI solutions. Moreover, they are well-positioned to contribute to the field by co-developing evaluation frameworks, participating in real-world testing scenarios, and helping shape domain-specific QA metrics that reflect the operational complexity of RE in practice. Their involvement is critical for transforming academic tools into trustworthy accountable systems.

## 5.4 | RQ4: Challenges and Future Directions

### 5.4.1 | Issue Interdependencies and Implications

Despite the considerable potential of GenAI in RE, our systematic analysis of the current research landscape revealed critical limitations that require a thorough investigation. By conducting a comprehensive review of 238 papers, we identified and hierarchically categorized 10 key challenges based on their prevalence and interdependencies. Based on the results of the analysis, which was supported by the correlation matrix shown in Figure 13, we established three distinct tiers of challenges based on their significance and impact on contemporary research. This tiered categorization provides a structured framework for understanding the complex interplay between promising opportunities and substantive challenges in the application of GenAI in RE.

Through this established framework, we ensured a comprehensive and systematic examination of the multifaceted nature of LLMs in the context of RE. The key challenges and limitations are as follows:

#### 5.4.1.1 | *Core Challenges in Current RE Practices (High Prevalence* > 50%*)*

- **Reproducibility (66.8%):** Reproducibility was identified as the most critical challenge in LLM-based RE research, with 66.8% of the reviewed studies explicitly addressing this issue. This high prevalence underscores the centrality of reproducibility for establishing stakeholder confidence, especially in requirement generation and validation scenarios where trust and reliability are paramount.

  In the context of RE, ensuring that GenAI systems consistently produce stable and verifiable results is essential but difficult. As highlighted by prior studies (Belz et al. (2021)), the inherent stochasticity of the LLM inference, coupled with the extreme sensitivity to seemingly minor parameter shifts (e.g., temperature and prompt phrasing), complicates validation in mission-critical tasks. These difficulties are magnified when commercial black-box LLMs where users often lack access to model internals or consistent API behaviors are employed, which makes the reliable regeneration of prior outputs nearly impossible.

  Although some studies acknowledge these challenges, few provide standardized solutions. Most evaluations are under specified, lacking the detailed documentation of parameter settings, prompt versions, or API states. This gap suggests an urgent need for community-level standardization, particularly regarding evaluation protocols, reproducibility reporting practices, and parameter disclosure guidelines, to enable cumulative knowledge building and enhance the robustness of GenAI adoption in RE tasks.

- **Hallucinations (63.4%):** Hallucinations are a core reliability issue of GenAI applications in RE, with 63.4% of the surveyed studies identifying it as a major concern, second in prevalence only to reproducibility. These hallucinations, often manifesting as input, context, or fact-conflicting outputs, pose substantial risks in RE, where the precision and traceability of requirements are essential for downstream system design and stakeholder alignment.

  Although various mitigation strategies ranging from data curation and reinforcement learning to external knowledge integration (Zhang et al. (2023), Maynez et al. (2020)), their adaptation and effectiveness within RE tasks remain largely underexplored. The specificity of RE, with its domain-critical and context-sensitive language, presents unique challenges that generic hallucination mitigation methods may not adequately address.

  Future research must prioritize the development of RE-specific countermeasures. These may include (1) automated evaluation metrics tailored to RE artifacts and semantics, (2) the adoption of structured multi-agent interaction frameworks for the cross verification of generated requirements, and (3) the use of uncertainty estimation methods to assess the confidence levels of generated outputs. Importantly, these efforts should be mindful of the trade-offs involved. Reducing hallucinations must not compromise the generative capacity or flexibility of the model, especially for complex or ambiguous stakeholder inputs. Addressing hallucinations in RE thus demands a delicate balance between control, accuracy, and generative capability within practical implementation scenarios.

- **Interpretability and explainability (57.1%):** Interpretability is the most prominent concern of GenAI applications in RE and is discussed in 57.1% of the reviewed studies. As GenAI systems increasingly automate tasks such as requirement analysis, elicitation, and classification, understanding the underlying decision-making processes becomes critical, particularly in high-stake domains such as healthcare and law. However, the intrinsic opacity of LLMs challenges efforts to provide transparent rationales behind the generated outcomes, which introduces risks related to validation, traceability, and accountability.

  Our analysis confirms that this interpretability limitation markedly affects stakeholder trust and poses a key barrier hindering the integration of LLMs into rigorous RE practices, as noted in prior studies (Zhao et al. (2023), Huang et al. (2023)). In particular, the complexity of how these models leverage pretrained knowledge and perform in-context adaptation remains poorly understood in RE contexts. Existing explainability techniques, although well studied in NLP and ML, have not been sufficiently adapted to capture the domain-specific semantics and stakeholder dynamics of RE tasks. Thus, the field urgently requires RE-specific interpretability frameworks that not only offer post hoc rationalizations but also support iterative collaboration, model traceability, and stakeholder validation across RE phases.

- **Controllability (50.8%):** Controllability is a frequently cited concern in LLM-driven RE and is identified in 50.8% of the reviewed studies. Unlike general-purpose NLP tasks, RE workflows often demand outputs conforming to strict domain conventions, stakeholder expectations, and regulatory constraints. However, current GenAI systems offer only limited mechanisms for precise behavioral steering, posing risks in environments where traceability, compliance, and alignment are nonnegotiable.

Although some studies have explored prompt engineering as a lightweight control mechanism, this approach alone is insufficient for aligning model behavior with nuanced organizational or project-specific requirements (Chong (2017), Yampolskiy (2020)). Particularly in safety-critical or highly dynamic RE contexts, the lack of fine-grained control over LLM outputs impedes consistent task execution and undermines stakeholder trust.

Our analysis reveals a research gap between the existing model control techniques often designed for general tasks and the granularity of control needed in RE scenarios. Moving forward, one needs to investigate structured control strategies such as constrained decoding, rule-based post-processing, modular pipelines, or function-calling interfaces allowing RE practitioners to inject formal constraints or dynamic feedback. Ultimately, bridging this gap will require designing controllability frameworks that are not only technically effective but also aligned with the iterative multistakeholder nature of RE.

#### 5.4.1.2 | *Evolving Concerns in RE Implementation (Prevalence = 20%-50%)*

- **Computational and economic cost (48.7%):** The computational and economic burden of deploying LLMs in RE contexts is an increasingly prominent issue, discussed in 48.7% of the reviewed studies. Unlike traditional RE automation tools, GenAI systems, especially large-scale pretrained models, entail substantial infrastructure demands, including high-performance computing, extensive memory bandwidth, and energy consumption. These requirements directly translate into operational costs and environmental impacts, which are rarely acknowledged in current RE-focused GenAI research.

  Our review reveals a notable oversight, showing that despite the resource intensity of LLM training and inference, many studies do not systematically evaluate the cost-effectiveness or sustainability of their proposed approaches. As noted in prior work (Strubell et al. (2019), Bender et al. (2021)), the centralization of GenAI capabilities in a handful of well-funded industrial actors raises further concerns regarding accessibility and equity in research. The current trajectory risks exacerbating disparities between organizations that can afford to experiment with GenAI at scale and those constrained by limited computational budgets.

  This situation calls for a research agenda focused on efficiency-aware GenAI-for-RE. Potential directions include the use of smaller task-tuned models (e.g., distilled or instruction-tuned LLMs), parameter-efficient fine-tuning (e.g., LoRA, QLoRA), and adaptive architectures that scale with task complexity. In addition, life-cycle cost analyses and carbon impact assessments should become standard components of RE system evaluations. Without such considerations, the integration of GenAI into RE will remain unsustainable economically and ecologically.

- **Bias and fairness (28.2%):** Bias and fairness remain underaddressed challenges in GenAI-supported RE and are acknowledged in only 28.2% of the surveyed studies. Despite this low coverage, the implications of algorithmic bias in RE are profound, particularly in the early phases such as requirement elicitation and analysis, where biased outputs can shape system specifications with far-reaching consequences.

  Our analysis reveals that current research primarily emphasizes functional performance, often at the expense of fairness. This oversight is concerning in high-stakes RE applications—such as public services, hiring systems, or law enforcement—where unmitigated biases can reinforce systemic inequities (Blodgett et al. (2020), Holstein et al. (2019)). Such risks are often subtle and embedded in domain-specific language patterns or stakeholder representations that pretrained models fail to contextualize appropriately.

  Although some studies have explored fairness-aware modeling and mitigation strategies (Bommasani et al. (2022)), they remain largely fragmented and generic and are rarely tailored to RE-specific workflows. Moving forward, a more principled approach that includes rigorous bias audits across RE phases, the integration of demographic-aware prompt engineering, and the curation of representative and inclusive requirement datasets is needed. These strategies will be essential not only for safeguarding ethical compliance, but also building GenAI-driven RE tools that are context-sensitive, trustworthy, and aligned with diverse user values.

- **Security and privacy (26.1%):** As LLMs increasingly permeate RE workflows, security and privacy emerge as foundational but underexplored concerns. Given that RE tasks often involve processing sensitive business logic, contractual obligations, and proprietary system requirements, overlooking these aspects can introduce notable operational and ethical risks.

  Although some works acknowledge security as an abstract quality requirement, few systematically exmine how GenAI systems can inadvertently expose or mishandle confidential requirement data. In adversarial scenarios, malicious actors may exploit model vulnerabilities to extract sensitive information, manipulate outputs, or even launch prompt injection attacks targeting mission-critical RE tasks (Carlini et al. (2021), Li et al. (2018)). These risks are further compounded when proprietary LLMs are used in black-box deployments without the guarantees of input isolation or data handling transparency.

  To mitigate these threats, future research must prioritize the development of RE-specific security protocols and privacy-preserving mechanisms. Potential directions include encryption-aware prompt handling, access-controlled LLM pipelines, the federated fine-tuning of local requirement data, and robust audit trails for traceability. Ultimately, ensuring that GenAI systems can uphold data integrity and confidentiality under real-world deployment conditions is not merely a technical necessity, but a prerequisite for responsible adoption in RE practice.

- **Ethical and regulatory concerns (23.1%):** Ethical and regulatory challenges represent a pressing but insufficiently addressed concern in the integration of GenAI into RE. This is a critical oversight, given that RE lies at the foundation of system behavior and stakeholder alignment domains, where unintended consequences from GenAI outputs can propagate downstream into deployed systems.

  The ethical risks associated with LLMs in RE are multifaceted, including the generation of biased or harmful requirement content, amplification of misinformation, and misinterpretation of stakeholder intentions (Whittaker et al. (2018), Jobin et al. (2019)). These concerns are particularly acute in sociotechnical systems where requirements touch upon legal, social, or equity-sensitive domains. However, current research rarely addresses the alignment of GenAI-generated requirements with ethical norms or compliance obligations.

  Moving forward, one should establish RE-specific ethical governance structures, e.g., develop transparent prompt logging and ethical review protocols for GenAI use in stakeholder-facing tasks and ensure the traceable provenance of automatically generated requirements. Importantly, these efforts should not be limited to academia but require active coordination among researchers, industry practitioners, legal experts, and policymakers to ensure that GenAI in RE evolves in a direction that is not only technically robust but also socially responsible.

#### 5.4.1.3 | *Emerging Considerations in RE Development (Low Prevalence < 20%)*

- **Real-time processing (18.5%):** Despite the increasing expectation for LLMs to support real-time interactions, such as dynamic dialogue management, adaptive requirement updates, and decision-making in volatile environments—only 18.5% of reviewed studies address this capability in the context of RE. This finding reveals a critical limitation in the current landscape, which remains largely focused on static or offline tasks such as requirement elicitation, classification, and specification.

  Real-time responsiveness introduces a distinct set of challenges for the use of GenAI in RE, e.g., latency constraints, system consistency under streaming inputs, and the need for contextual awareness as requirements evolve in tandem with stakeholder input or project changes. However, few studies explicitly consider how GenAI systems can accommodate such dynamism. As highlighted in a recent research (Team (2024)), the lack of RE frameworks designed for continuous input integration or temporal requirement validation remains a key technical gap.

  To advance toward operationally viable GenAI-for-RE, future research must explore architectures and pipelines that support continuous monitoring, incremental inference, and feedback-driven adaptation. This exploration can involve hybrid LLM microservice systems, memory-augmented models, or real-time fine-tuning mechanisms tailored to RE domains. Bridging the current gap between static GenAI pipelines and real-time RE needs is essential to ensure that GenAI can play a meaningful role in agile, iterative, and time-sensitive SE practices.

- **Authorship and copyright (8.8%):** Despite the growing application of GenAI in RE, intellectual property (IP) concerns remain markedly underexplored, with only 8.8% of the reviewed studies addressing authorship and copyright issues. This is especially problematic given the increasing fluency of LLMs in generating structured requirements, design rationale, and documentation artifacts that are typically governed by clear ownership conventions in traditional SE workflows.

  As highlighted by (Deltorn and Macrez (2018)), the legal status of AI-generated artifacts in RE is ambiguous. Uncertainties persist over whether outputs from GenAI models qualify for protection under existing copyright laws, and if so, who should be regarded as the rightful author the user, model provider, or organization deploying the tool. These ambiguities not only pose compliance risks for RE practitioners, but also complicate the reuse, licensing, and traceability of requirement assets throughout the project lifecycle.

  Although several studies have explored these questions (Ihalainen (2018)), the field lacks a structured discourse on how GenAI fits into established IP frameworks. Future interdisciplinary research must bridge this gap by articulating context-sensitive legal standards, exploring attribution models tailored for collaborative human–AI authorship, and formulating governance principles that balance innovation with accountability. Establishing such frameworks will be crucial for legitimizing GenAI outputs in RE, particularly in regulated or contractual environments where provenance and legal standing of requirements are nonnegotiable.

### 5.4.2 | Correlation Analysis and Impact

The analysis of the correlation patterns in the challenges associated with LLMs revealed a hierarchical structure of interdependencies in RE applications. The updated correlation matrix illustrated in Figure 13 demonstrates three distinct tiers of challenges, each characterized by varying degrees of prevalence and interrelation. This tiered structure suggests that the effective mitigation of these challenges requires a strategic layered approach acknowledging their standalone impact and cooccurrence dynamics. The corresponding matrix quantifies two core dimensions, namely the individual prevalence of each challenge (diagonal elements) and pairwise cooccurrence frequencies (off-diagonal elements). The darker blue tones in the heatmap indicate stronger correlations. The following formulations are used.

$$\text{Individual issue percentage} = \frac{\text{Papers discussing the issue}}{\text{Total papers}} \times 100\%, \tag{1}$$

$$\text{Cooccurrence percentage} = \frac{\text{Papers discussing both issues}}{\text{Total papers}} \times 100\%. \tag{2}$$

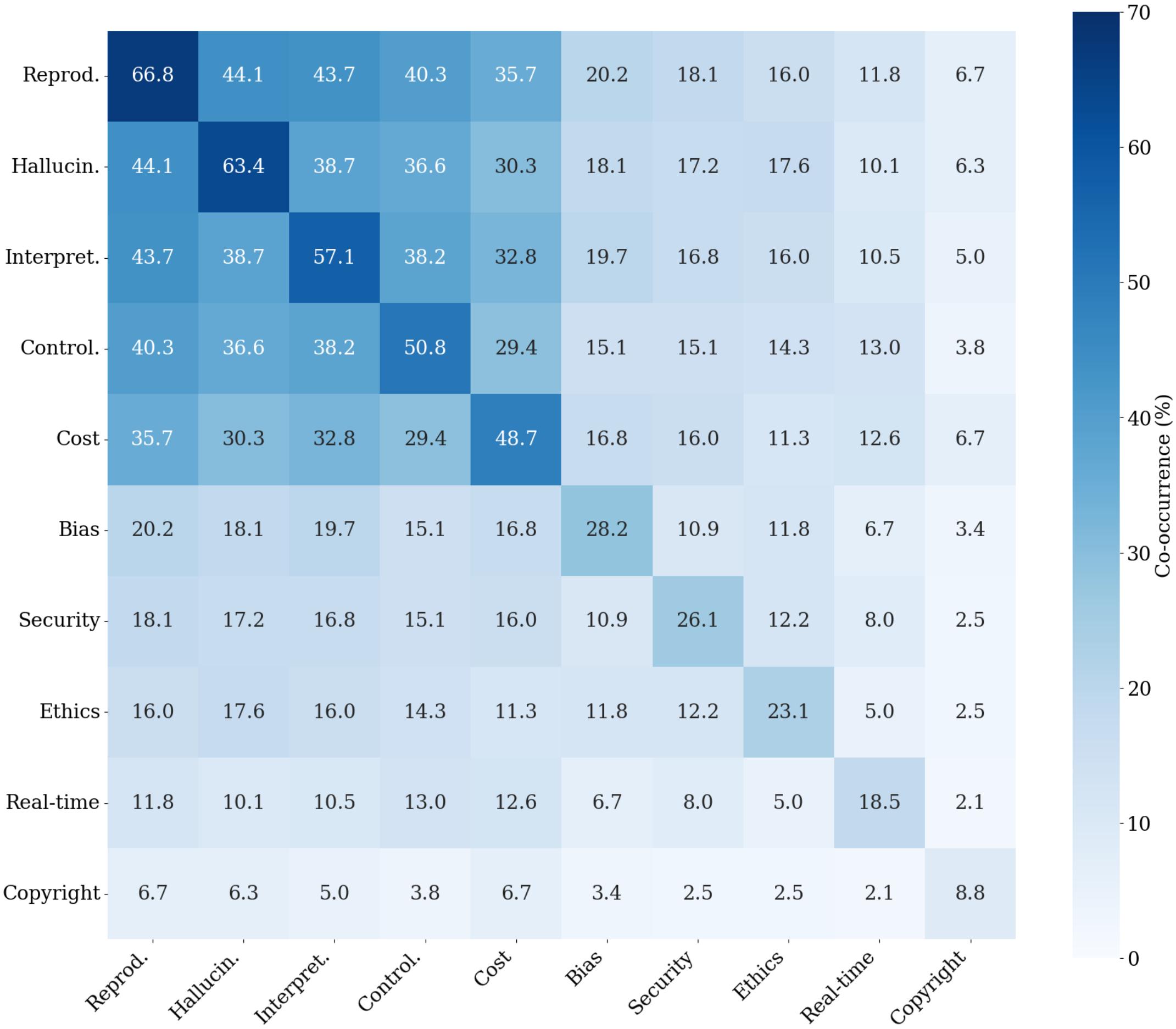

**FIGURE 13** Correlations among the LLM issues reported in literature on RE (%).

For example, the diagonal value for reproducibility is 66.8%, indicating that it is the most frequently discussed challenge in the dataset. The off-diagonal cell between reproducibility and interpretability (41.0%) reveals that a substantial portion of the examined studies recognizes the interconnectedness between these two concerns. Such correlations offer insights into how the technical challenges of GenAI use in RE often cluster in practice, both conceptually and empirically.

The strongest triadic relationship is composed of reproducibility (66.8%), hallucination control (63.4%), and interpretability (57.1%), surpassing previous triads by prevalence and mutual linkage. These three challenges exhibit a high cooccurrence rate (35%), reflecting their tight integration in the RE contexts. This finding underscores that improving reproducibility often necessitates increased transparency (interpretability) and the mitigation of misleading or unfaithful outputs (hallucinations). In particular, the 41.0% cooccurrence between reproducibility and interpretability indicates that the mutual reinforcement loop involving unclear decision processes complicates verification, while inconsistent outputs undermine interpretive clarity.

The second tier of influential yet more context-dependent challenges includes controllability (50.8%) and computational cost (48.7%), both of which show moderate correlations (25%–30%) with the top triad. These challenges often emerge when transitioning from proof-of-concept (PoC) models to real-world deployment. Controllability, for example, is not only a technical constraint but also an operational one, which is closely tied to the reproducibility and customizability of model behavior. Similarly, computational cost cuts across technical feasibility and broader accessibility, acting as a bottleneck for implementing more sophisticated solutions such as RAG or multimodel integration.

Low-prevalence issues, including bias (28.2%), security (26.1%), and ethics (23.1%), form a cluster of governance and accountability concerns. Their intercorrelations (8%-16%) suggest shared foundations in value-sensitive design and stakeholder trust, albeit with limited direct influence on technical challenges. Copyright (8.8%) remains the most isolated issue, with cooccurrence rates consistently below 5% suggesting that IP concerns have yet to be integrated into mainstream RE discussions, despite their rising relevance as LLM-generated artifacts in fidelity and adoption.

Overall, the correlation structure affirms that the challenges in GenAI use in RE are not isolated phenomena but interdependent dynamics shaping the effectiveness and trustworthiness of real-world systems. Addressing these challenges demands a system-thinking approach that bridges technical precision, resource feasibility, and ethical accountability to create sustainable GenAI solutions for RE.

**Main Findings for RQ4**

Our correlation analysis reveals that the challenges in applying GenAI in RE are not isolated but form a highly interdependent structure. At the center of this structure lies a core triad of **reproducibility (66.8%)**, **hallucination control (63.4%)**, and **interpretability (57.1%)**. These issues exhibit high prevalence and strong cooccurrence, thus jointly defining the operational and epistemic limits of LLM-based RE. Their entanglement suggests that any intervention targeting one must account for ripple effects on the others, making piecemeal solutions insufficient. A second layer of deployment-facing challenges—**controllability (50.8%)** and **computational cost (48.7%)**—serves as a strategic hinge between conceptual feasibility and real-world implementation. Governance-related concerns such as **bias**, **security**, and **ethics** form a loosely correlated cluster, signaling emerging but underaddressed sociotechnical risks. **Copyright (8.8%)** stands as a conceptually isolated issue, hinting at a widening gap between legal frameworks and practices of GenAI use in RE. Overall, our findings emphasize that effective GenAI integration in RE demands holistic multilayered strategies that bridge technical robustness, operational viability, and ethical accountability.

**Takeaway Message 4**

**For researchers:** Correlation analysis reveals that the challenges of using GenAI in RE are deeply interconnected. The triad of core producibility, hallucination control, and interpretability defines a foundational design space for future research. Addressing one dimension in isolation may prove ineffective without considering its effects on the others. Moreover, controllability and cost serve as deployment-sensitive pivots, requiring research that combines performance, efficiency, and traceability. The emergence of governance-related clusters indicates the need for value-aligned GenAI systems and calls for interdisciplinary collaboration across SE, human–computer interaction, ethics, and legal studies. Finally, the neglect of copyright concerns signals an opportunity for pioneering research into ownership, attribution, and compliance frameworks tailored for AI-generated RE artifacts.

**For practitioners:** Practitioners should interpret these challenges not as isolated technical issues but as interconnected risk dimensions. Investing in transparency and reproducibility will strengthen stakeholder trust and auditability, especially in mission-critical domains. However, successful deployment also depends on controlling the computational overhead and tailoring system outputs to project-specific constraints. Beyond system-level concerns, practitioners must remain alert to emerging governance liabilities, including bias propagation, data leakage, and IP ambiguities. These findings underscore the importance of building compliance-aware and ethically aligned RE pipelines that not only meet functional requirements but also align with broader social, legal, and organizational norms.

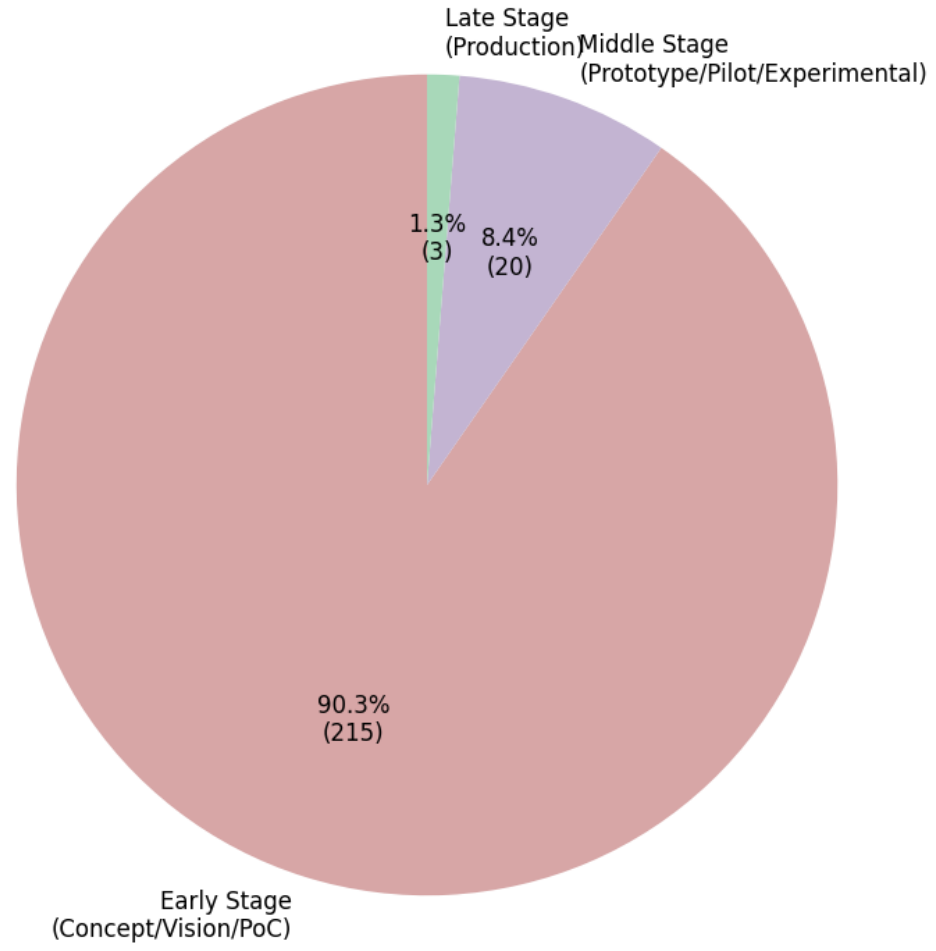

**FIGURE 14** Industrial adoption stages of GenAI use in RE (N=238).

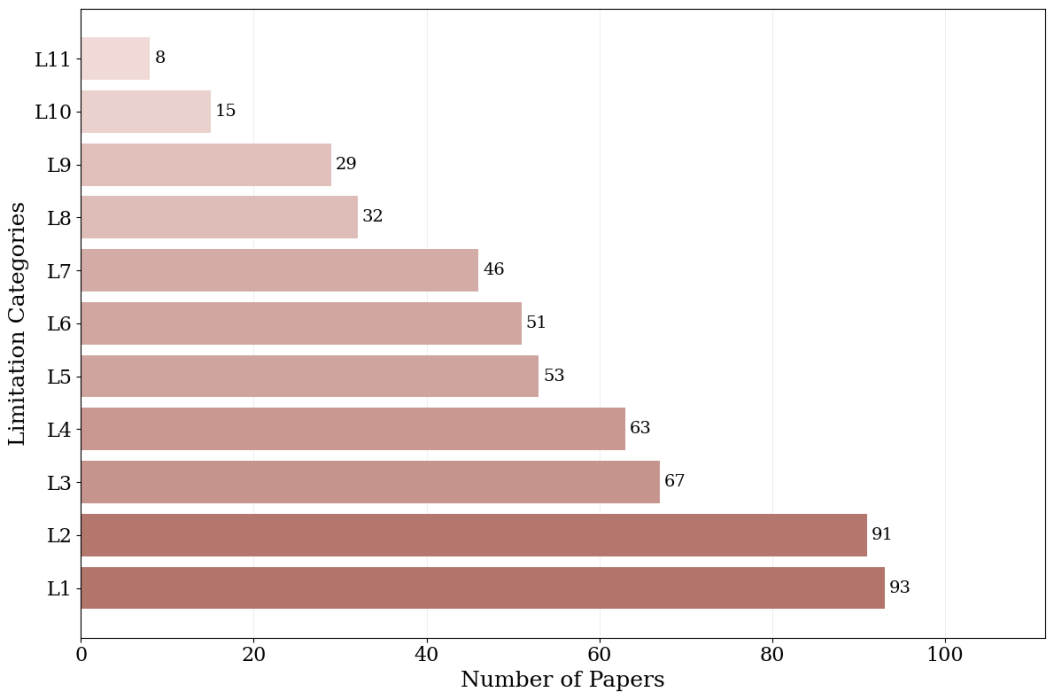

**FIGURE 15** Distribution of the major limitation categories (N=238). L1 = generalizability and domain adaptation, L2 = data quality and availability, L3 = evaluation and validation, L4 = human involvement and manual work, L5 = prompt engineering and configuration, L6 = scalability and resource constraints, L7 = model performance and reliability, L8 = model behavior issues, L9 = real-world deployment, L10 = security and privacy, L11 = transparency and control.

## 5.5 | RQ5: Industrial Adoption Status and Systemic Limitations

Despite the growing enthusiasm surrounding GenAI use in RE, our systematic review reveals that its practical adoption remains overwhelmingly nascent. As illustrated in Figure 14, most studies (90.3%) correspond to an early stage of development, focusing primarily on conceptual discussions, vision papers, or PoC prototypes. Only 8.4% report efforts at the prototype or experimental deployment level, and a mere 1.3% reach production-grade implementations. This striking imbalance underscores the fundamental gap between academic exploration and industrial integration, suggesting that most current GenAI–RE solutions are still far from being operationalized in real-world settings.

Our investigation of explicit limitation tagging further highlights the immaturity of the field. Numerous papers (29.6%) directly acknowledge barriers or limitations to implementation, while only 4.3% reference open-source releases, and 10.7% explicitly state that their work is not in production. These annotations highlight the experimental nature of most research and lack of translational pathways from lab-based validation to field-level deployment.

To better understand the bottlenecks impeding industrial readiness, we systematically categorized the reported limitations into 11 classes (Figure 15). The most frequently cited constraints include generalizability and domain adaptation (39.9%) and data quality and availability (39.1%), indicating that foundational input-related challenges remain unsolved. These are followed by concerns regarding evaluation methodologies (28.8%) and manual involvement (27.0%), both of which reflect a methodological underdevelopment and lack of scalable validation pipelines. Notably, 47.2% of the selected studies are associated with three or more limitation categories, pointing to the systemic multilayered nature of GenAI integration in RE.

**Main Findings for RQ5**

Our review reveals a critical disconnect between academic innovation and real-world adoption in the use of GenAI in RE. Despite the rich conceptual advancements, >90% of the reviewed studies remain at the vision or prototype level, with production-grade applications being exceedingly rare. The identified limitations cluster around input fragility (e.g., data quality and generalizability), methodological shortcomings (e.g., evaluation practices), and implementation complexity (e.g., scalability and prompt engineering), suggesting that current approaches face systemic rather than isolated obstacles. These findings reflect a field that, while innovative, is still in the early stages of operational maturity.

**Takeaway Message 5**

**For researchers:** The limited real-world deployment of GenAI in RE should be viewed not as a weakness but as a rich opportunity space for translational research. Addressing core limitations, such as generalizability, data availability, and scalable evaluation methods, requires interdisciplinary collaboration, access to industry-grade datasets, and the development of domain-specific benchmarks. Moreover, advancing toward production readiness demands an architectural shift toward modular, robust, and explainable GenAI–RE systems that can operate reliably across heterogeneous environments.

**For practitioners:** The findings serve as a cautionary note regarding the immediate readiness of GenAI tools for production RE environments. Although the potential is promising, current research lacks the validation, reliability, and transparency needed for enterprise-level integration. Practitioners are advised to engage cautiously, prioritizing pilot deployments and collaboration with academic partners to shape fit-for-purpose tools. Investment in interpretability, traceability, and security should be emphasized when evaluating GenAI-based RE tools for operational use.

# 6 | THREATS TO VALIDITY

In this SLR, we conducted a comprehensive assessment of potential threats to the validity of our findings, considering the internal, external, and construct validity aspects. By critically examining such threats, we aimed to provide a transparent and rigorous evaluation of the limitations and strategies used to mitigate the threats. Internal validity, which pertains to the robustness and integrity of the research design and execution, is subject to two primary threats. First, the inherent risk of inadvertently omitting relevant papers remains despite the implementation of a systematic literature search and screening method. This threat emerges from the possibility of papers being indexed in databases that are not covered by the search strategy or papers that use alternative terminologies that were not captured by search strings. To mitigate this threat, we conducted a comprehensive search across multiple reputable databases and adopted a meticulously crafted set of inclusion and exclusion criteria to ensure the identification of pertinent literature. Second, the process of data extraction and analysis is susceptible to subjective judgment and potential bias introduced by individual researchers. To address this concern, we adopted a standardized data extraction framework to maintain consistency and followed a rigorous cross-checking procedure that involved three independent researchers to minimize bias and ensure the reliability of extracted data.

External validity, which pertains to the generalizability and applicability of the findings to wider contexts, is prone to two primary threats that may circumscribe the extent to which the results can be extrapolated beyond the specific research setting. The temporal scope of the study, which focused on literature published from 2019 to May 31, 2025, may only partially capture the comprehensive state of GenAI applications in RE. This limitation is particularly relevant given the rapid evolution and proliferation of GenAI technologies in recent years. Consequently, the findings may not entirely reflect the most recent advancements and innovations in the field. In addition, the generalizability of the conclusions of this study may be limited by the specific characteristics and contexts of the included papers, such as the domain of application, scale of projects, and cultural or organizational settings in which the research was conducted. These limitations underscore the importance of interpreting the findings with caution and considering specific contextual factors when applying the findings to different scenarios. Construct validity, which relates to the definition and measurement of research concepts, presents challenges inherent to GenAI application in RE. The rapid evolution of GenAI technologies and their application in RE necessitates the continuous refinement and adaptation of evaluation frameworks to ensure the relevance and effectiveness of GenAI technologies. The analysis framework proposed in this study, although based on a comprehensive examination of the literature and expert consultation, may not exhaustively encompass all aspects and dimensions of GenAI applications in RE. This limitation highlights the need for continued research and discourse to identify and incorporate additional factors that can influence the effectiveness and impact of these technologies in practice.

To mitigate these threats to result validity, we employed a multipronged approach. First, we employed a rigorous and systematic literature search and screening process that used multiple databases and adopted well-defined data inclusion and exclusion criteria to minimize the risk of omitting relevant papers. Second, multiple researchers conducted cross-checking to ensure the reliability and consistency of data extraction and analysis, reducing the influence of individual bias. Third, the temporal and technological limitations of this review were explicitly acknowledged, emphasizing the need for a continued update and expansion of the review as the field progresses. Fourth, the use of widely accepted QA criteria was complemented by a critical reflection on their potential limitations and the recognition of the need for tailored evaluation frameworks specific to GenAI applications in RE. Finally, the classification framework was iteratively refined through a combination of literature analysis and expert consultation, which aimed to capture the most salient aspects of the field while acknowledging the potential for further enhancement.

# 7 | ROADMAP FOR ADVANCING GENAI APPLICATIONS IN RE

To enable the meaningful, sustainable, and trustworthy adoption of GenAI in RE, we propose a multiphase research roadmap synthesized from our cross-sectional findings:

**Reinforcement of evaluation foundations:** The dominance of early-stage work (Figure 14) underscores the pressing need to strengthen the evaluation infrastructure. Standardized benchmarks, RE-specific metrics, and reproducibility protocols should be prioritized. Given the lack of real-world deployment (only 1.3% of studies reach production), establishing unified evaluation pipelines is essential for aligning academic exploration with industrial constraints.

**Governance-aware development:** Governance-related issues, including fairness, interpretability, ethics, and privacy, remain underaddressed but are becoming increasingly important. Our cooccurrence analysis (Figure 13) revealed that these challenges are often clustered, indicating that siloed solutions may be insufficient. Future development should integrate ethical auditing, fairness constraints, and stakeholder-centered validation as parts of designing GenAI systems for RE.

**Toward scalable and context-aware deployment:** To transition from PoC to practice, scalable and cost-effective deployments must be enabled, as exemplified by the use of modular architectures, fine-tuning strategies (e.g., LoRA and PEFT), and system-level solutions (e.g., RAG) to reduce hallucinations and improve control. Hybrid systems supporting real-time updates and traceability will be instrumental in bridging academic prototypes with industrial-grade systems.

**Industrial-scale maturity and standardization:** As shown in our maturity analysis Figure 14, most works (90%) remain in conceptual stages. For broader adoption, community-driven toolkits, open-source benchmarks, and legal frameworks (e.g., authorship and copyright governance) must mature. Collaboration with standard bodies and industrial consortia are critical for establishing certification-ready GenAI workflows for RE.

This roadmap emphasizes the dual trajectory of advancing the technical robustness of GenAI while ensuring ethical, legal, and social readiness for integration into the RE lifecycle.

# 8 | CONCLUSION

This SLR presents a comprehensive and fine-grained investigation of GenAI applications in RE, synthesizing insights from 238 peer-reviewed papers published between 2019 and May 2025 and identifying key trends, methodological patterns, and structural gaps shaping the trajectory of this rapidly evolving research field.

A concentration of efforts is observed in the early RE phases, such as elicitation (22.1%) and analysis (30.0%), with later stages such as requirement management (6.8%) remaining substantially underrepresented. This functional imbalance signals the need for a broader exploration of the capabilities of GenAI across the entire RE lifecycle. Additionally, the dominance of GPT-series models (67.7%) suggests heavy reliance on a single LLM lineage, highlighting the need for diversification and contextual adaptation.

The most prevalent challenges, namely reproducibility (66.8%), hallucinations (63.4%), and interpretability (57.1%), form a tightly interlinked triad affecting trust, consistency, and accountability in RE practices. Importantly, our correlation analysis reveals that addressing these challenges in isolation is insufficient and indicates the need for holistic system level strategies. In contrast, concerns such as bias, ethics, and real-time adaptability, though less frequently reported, are critical for ensuring the socially responsible and operationally resilient deployment of GenAI.

The analysis of prompt engineering techniques and learning paradigms reveals a trend toward instruction-based prompts (62.2%) and few-shot learning setups (43.6%). These results reflect the ease of adoption and evolving sophistication of GenAI-RE toolchains. However, much of this experimentation remains confined to prototypes. Our industrial readiness review shows that most studies( >90%) remain in early stage conceptual or PoC phases, with only 1.3% reaching production-level integration. Limitations such as generalizability, data quality, and scalability appear frequently and in combination, reflecting deep-seated systemic bottlenecks rather than isolated technical flaws.

Evaluation practices also reveal considerable maturity gaps. Our synthesis of the four key evaluation dimensions shows that although many studies report tool or dataset usage, actual public availability remains limited. Benchmarking remains fragmented, while evaluation metrics heavily favor surface-level indicators such as precision and recall. Triangulated human-centered validation—essential for the sociotechnical nature of RE is rarely observed. This fragmentation suggests that current research lacks the transparency and empirical rigor needed to generalize findings or inform industrial deployment.

The successful adoption of GenAI in RE hinges on technical robustness, methodological maturity, and governance integration. Technical robustness entails moving beyond single-model paradigms toward architectures that balance performance, interpretability, and controllability. Methodological maturity requires the standardization of evaluation frameworks and adoption of hybrid empirical designs. Governance integration calls for the codevelopment of ethical, legal, and security protocols alongside algorithmic innovation.

In conclusion, although GenAI holds transformative promise for RE, its full realization demands coordinated efforts across research, industry, and policymaking. This SLR offers not only a critical synthesis of current practices but also a structured roadmap for enabling responsible, scalable, and context-aware adoption of GenAI in RE. Future work must build on these insights to ensure that the field evolves from exploratory promise to dependable real-world impact.

## ACKNOWLEDGMENTS

This work was supported by JST SPRING (grant number JPMJSP2128) and the JST-Mirai program (grant number JPMJMI20B8).

# DATA AVAILABILITY

The link to the data is available in the manuscript.

## FINANCIAL DISCLOSURE

None.

## CONFLICT OF INTEREST

The authors declare no potential conflicts of interest.